\documentclass{emulateapj}
\usepackage{epsfig}

\newcommand{\beq}{\begin{equation}}
\newcommand{\eeq}{\end{equation}}
\def\be{\begin{equation}}
\def\ee{\end{equation}}
\def\bea{\begin{eqnarray}}
\def\eea{\end{eqnarray}}



\def \logTd6 {\hbox{log$( T/6 \kev)$} }

\def\myputfigure#1#2#3#4#5%
{\vskip#5pt\makebox[0pt]{\hskip#2in
\includegraphics[width=#3\textwidth]{#1}}\vskip#4pt\hfill}

\def \arcmin     { ^{\prime} }
\def \arcsec    {^{\prime\prime}}

\def \kms            {~{\rm km~s}^{-1}}

\def \etal      {et al.}

\def \kev       {{\rm\ keV}}

\def \hMpc      {h^{-1}{\rm\ Mpc}}
\def \hkpc      {h^{-1}{\rm\ kpc}}

\newcommand{\mnhi}{N_{\rm HI}}

\newcommand{\lya}{Ly$\alpha$}

\input epsf
\bibpunct{(}{)}{;}{a}{}{,}

\begin{document}

\lefthead{QUASARS PROBING QUASARS}
\righthead{HENNAWI \& PROCHASKA}

\title{Quasars Probing Quasars II: The Anisotropic Clustering of
  Optically Thick Absorbers around Quasars}

\author{Joseph F. Hennawi\altaffilmark{1,2} \& Jason X. Prochaska\altaffilmark{3}}
 
\altaffiltext{1}{Department of Astronomy, University of California
  Berkeley, Berkeley, CA 94720; joeh@berkeley.edu}
\altaffiltext{2}{Hubble Fellow}
\altaffiltext{3}{Department of Astronomy and Astrophysics, 
  UCO/Lick Observatory; University of California, 1156 High Street, Santa Cruz, 
  CA 95064; xavier@ucolick.org}

\begin{abstract}
  With close pairs of quasars at different redshifts, a background
  quasar sightline can be used to study a foreground quasar's
  environment in \emph{absorption}. We used a sample of 17 Lyman limit
  systems with column density $\mnhi > 10^{19}~{\rm cm^{-2}}$ selected
  from 149 projected quasar pair sightlines, to investigate the
  clustering pattern of optically thick absorbers around luminous
  quasars at $z\sim 2.5$. Specifically, we measured the
  quasar-absorber correlation function in the transverse direction,
  and found a comoving correlation length of
  $r_0=9.2^{+1.5}_{-1.7}~\hMpc$ (comoving) assuming a power law
  correlation function, $\xi\propto r^{-\gamma}$, with $\gamma=1.6$.
  Applying this transverse clustering strength to the line-of-sight,
  would predict that $\sim 15-50\%$ of \emph{all} quasars should show
  a $\mnhi > 10^{19}~{\rm cm^{-2}}$ absorber within a velocity window
  of $\Delta v < 3000~\kms$. This overpredicts the number of absorbers
  along the line-of-sight by a large factor, providing compelling
  evidence that the clustering pattern of optically thick absorbers
  around quasars is highly anisotropic.  The most plausible
  explanation for the anisotropy is that the transverse direction is
  less likely to be illuminated by ionizing photons than the
  line-of-sight, and that absorbers along the line-of-sight are being
  photoevaporated.  A simple model for the photoevaporation of
  absorbers subject to the ionizing flux of a quasar is presented, and
  it is shown that absorbers with volume densities $n_{\rm H} \lesssim
  0.1~{\rm cm}^{-3}$ will be photoevaporated if they lie within $\sim
  1~{\rm Mpc}$ (proper) of a luminous quasar.  Using this simple
  model, we illustrate how comparisons of the transverse and
  line-of-sight clustering around quasars can ultimately be used to
  constrain the distribution of gas in optically thick absorption line
  systems.
\end{abstract}

\keywords{quasars: general -- intergalactic medium -- quasars: absorption lines -- cosmology: general -- surveys: observations}

\section{Introduction}
\label{sec:intro}

\begin{figure}
  \centerline{\epsfig{file=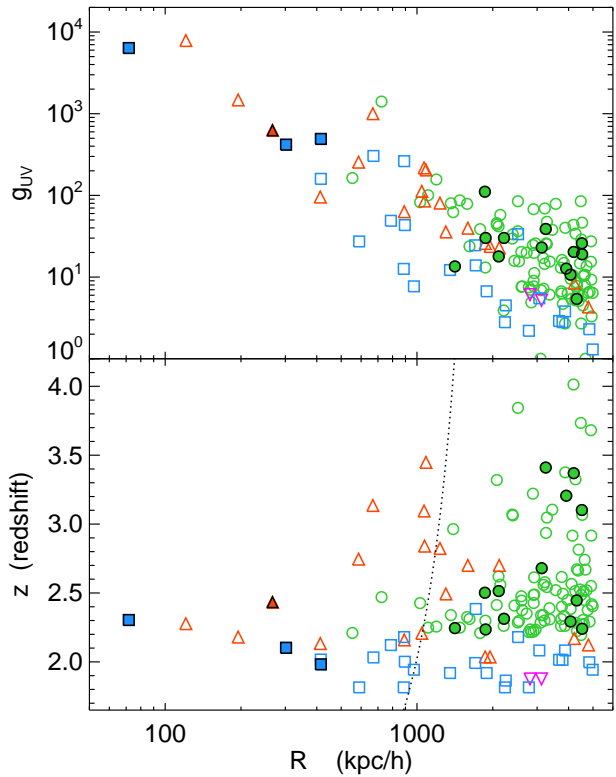,bb=100 80 600 680,
      width=0.50\textwidth}}
  \caption{Distribution of foreground quasar redshifts, transverse
    separations, and ionizing flux ratios probed by the background quasar
    sightlines.  The upper plot shows the ratio of the ionizing flux
    to the UV background $g_{\rm UV}$, note the general $R^{-2}$
    trend.  The lower plot shows foreground quasar redshift versus
    transverse comoving separations.  The (blue) squares have a Keck
    spectrum of the background quasar, (red) triangles have Gemini
    background spectra, (magenta) upside down triangles have MMT
    background spectra, and (green) circles have SDSS background
    spectra. Filled symbols outlined in black have a super-LLS within
    $|\Delta v| < 1500\kms$ at the foreground quasar redshift (see
    Table~\ref{table:slls}) and open symbols have no absorber.  The
    region to the left of the dotted line is excluded by the SDSS
    fiber collision limit of $\theta=55\arcsec$, which explains the
    paucity of SDSS background spectra there.  The Keck/Gemini/MMT
    spectra probe angular separations an order of magnitude smaller
    than the fiber collision limit, allowing us to study the
    foreground quasar environment down to down to $70~\hkpc$ where the
    ionizing flux is $\sim 10,000$ times the UV
    background.\label{fig:scatter}}
\end{figure}

Although optically thick absorption line systems, that is Lyman
Limit Systems (LLSs) and damped Lyman-$\alpha$ systems (DLAs), are
detected as the strongest absorption lines in quasar spectra, the two
types of objects, quasars and absorbers, play rather different roles
in the evolution of structure in the Universe.  The hard ultraviolet
radiation emitted by luminous quasars gives rise to the ambient
extragalactic ultraviolet (UV) background \citep[see
e.g.][]{HM96,Miralda03,Meiksin05} responsible for maintaining the low neutral
fraction of hydrogen ($\sim 10^{-6}$) in the intergalactic medium (IGM),
established during reionization. However, high column density
absorbers represent the rare locations where the neutral fractions are
much larger.  Gas clouds with column densities $\log N_{\rm HI}>17.2$
are optically thick to Lyman continuum ($\tau_{\rm LL} \gtrsim 1$)
photons, giving rise to a neutral interior self-shielded from the
extragalactic ionizing background.  In particular, the damped
Ly$\alpha$ systems dominate the neutral gas content of the Universe
\citep{phw05}, which is the primary reservoir for the formation of 
stars observed in local galaxies. 

One might expect optically thick absorbers to be absent at small
separations from luminous quasars.  For example, a quasar at $z=2.5$
with magnitude $r=19$, emits a flux of ionizing photons
that is 130 times higher than that of the extragalactic UV background
at an angular separation of $60\arcsec$, corresponding to a proper
distance of 340~$\hkpc$.   Indeed, the decrease in the number of
\emph{optically thin} absorption lines ($\log N_{\rm HI}<17.2$ hence
$\tau_{\rm LL}\lesssim 1$), in the vicinity of quasars, known as the
\emph{proximity effect} \citep{BDO88}, provides a measurement of the
UV background \citep{Scott00}. If Nature provides a nearby background
quasar sightline, one can also study the \emph{transverse proximity
  effect}, which is the expected decrease in absorption in a
\emph{background} quasar's Ly$\alpha$ forest, caused by the transverse
ionizing flux of a \emph{foreground} quasar. The transverse effect has
yet to be detected, in spite of many attempts \citep[][but see
Jakobsen \etal 2003]{Crotts89,DB91,FS95,LS01,Schirber04,Croft04}.

On the other hand, it has long been known that quasars are associated
with enhancements in the distribution of galaxies
\citep{BSG69,YG84,YG87,BC91,SBM00,BBW01,Serber06,Coil06}, although
these measurements of quasar galaxy clustering are mostly limited to
low redshifts $\lesssim 1.0$. Recently, \citet{AS05}, measured the
clustering of Lyman Break Galaxies (LBGs) around luminous quasars in
the redshift range ($2 \lesssim z \lesssim 3.5$), and found a best fit
correlation length of $r_0=4.7~\hMpc$ ($\gamma=1.6$), very similar to
the auto-correlation length of $z\sim 2-3$ LBGs
\citep{Adel03}. \citet{Cooke06} recently measured the clustering of
LBGs around DLAs and measured a best fit $r_0=2.9~\hMpc$ with
$\gamma=1.6$, but with large uncertainties \citep[see
also][]{Gawiser01,Bouchet04}.  If LBGs are clustered around quasars,
and LBGs are clustered around DLAs, might we expect optically thick
absorbers to be clustered around quasars?  This is especially
plausible in light of recent evidence that DLAs arise from a high
redshift galaxy population which are not unlike LBGs
\citep{Schaye01,Moller02}.

Clues to the clustering of optically thick absorbers around
quasars come from \emph{proximate DLAs}, which have absorber redshifts 
within $3000\kms$ of the emission redshift of the quasars \citep[see
e.g.][]{Moller98}. Recently, \citet{REB06} \citep[see
also][]{Ellison02}, compared the number density of proximate
DLAs per unit redshift to the average number density of DLAs in the
the Universe \citep{phw05}. They found that the abundance of
DLAs is enhanced by a factor of $\sim 2$ near quasars, which they
attributed to the clustering of DLA-galaxies around quasars.

Projected pairs of quasars with small angular separations ($\theta
\lesssim 5\arcmin$) but different redshifts, can also be used to study
the clustering of absorbers around quasars.  In this context,
clustering is manifest as an excess probability, above the cosmic
average, of detecting an absorber in the neighboring background quasar
spectrum, near the redshift of the foreground quasar.  In the first of
this series of four papers on optically thick absorbers near quasars
\citep[][henceforth QPQ1]{LLS1}, we used background quasar
sightlines to search for optically thick absorption in the vicinity of
foreground quasars: 149 projected quasar pairs were systematically
surveyed for Lyman limit Systems (LLSs) and Damped Ly$\alpha$ systems
(DLAs) in the vicinity of $1.8 < z < 4.0$ luminous foreground
quasars. A sample of 27 new absorbers were uncovered with transverse
separations $R < 5~\hMpc$ (comoving) from the foreground quasars, of
which 17 were super-LLSs with $\mnhi > 10^{19}~{\rm cm^{-2}}$.

The distribution of foreground quasar redshifts, transverse
separations, and ionizing flux ratios probed by the projected pair
sightlines studied in QPQ1 are illustrated in
Figure~\ref{fig:scatter}.  The filled symbols outlined in black
indicate the sightlines which have an absorption line system with
$\mnhi > 10^{19}~{\rm cm}^{-2}$ within a velocity interval of
$\left|\Delta v\right| = 1500~\kms$ of the foreground quasar redshift,
and the open symbols represent sightlines with no such absorber. The
line density of absorbers per unit redshift at $z\sim 2.5$ with column
densities $\mnhi > 10^{19}~{\rm cm}^{-2}$ is $dN\slash dz \simeq
0.8$\citep{Omeara06}, which implies that the expected number of
\emph{random} quasar-absorber coincidences within $|\Delta
v|=1500\kms$ is $\sim 3\%$.  The expected number of random
quasar-absorber associations in the sample of 149 sightlines is
$\langle N \rangle = 4.4$; whereas, 17 systems were discovered in QPQ1
-- compelling evidence for clustering.

This clustering is particularly conspicuous on small scales: out of 8
sightlines with $R < 500~\hkpc$, 4 quasar-absorber pairs were
discovered, which implies a small scale covering factor of $\sim
50\%$, whereas $\langle N\rangle = 0.18$ would have been expected at
random. This excess of $\sim 25$ over random may not come as a
surprise when one considers that galaxies are strongly clustered
around quasars on small scales. However, the light from \emph{every}
isolated quasar in the Universe traverses these same small scales on
the way to Earth.  Naively, we would then expect a comparable covering
factor of $\mnhi > 10^{19}~{\rm cm}^{-2}$ super-LLSs along the
line-of-sight (LOS) -- which is definitely not observed to be the
case!

How can we quantify the clustering of absorbers around quasars? How
do we compare the transverse clustering pattern to the number of
proximate absorbers observed along the LOS?  What can
the quasar-absorber clustering pattern teach us about quasars and
absorbers?  These questions are addressed in this second paper of the
series. We briefly review the QPQ1 survey of projected quasar pairs and
discuss the quasar-absorber sample in \S~\ref{sec:sample}.  A
formalism for quantifying the line-of-sight (LOS) and transverse
clustering of absorbers around quasars is presented in
\S~\ref{sec:formalism}.  In \S~\ref{sec:maxL} we introduce a maximum
likelihood technique for estimating the quasar-absorber correlation
function, which we use to measure the clustering of our
quasar-absorber sample in \S~\ref{sec:cluster}. Our measurement of the
transverse clustering is used to predict the expected number of
proximate absorption line systems which should be observed along the
line-of-sight in \S~\ref{sec:proximity}. In \S~\ref{sec:photo}, we
introduce a simple analytical model for the photoevaporation of
optically thick absorbers by quasars, which is used to illustrate how
comparisons of the line-of-sight and transverse clustering can
constrain the distribution of gas in optically thick absorbers. 
We summarize and discuss our results in \S~\ref{sec:conc}

Paper III of this series \citep{LLS3} investigates fluorescent
Ly$\alpha$ emission from our quasar-absorber pairs, and echelle
spectra of several of the quasar-LLS systems published here are
analyzed in Paper IV \citep{LLS4}.  Throughout this paper we use the
best fit WMAP (only) cosmological model of \citet{Spergel03}, with
$\Omega_m = 0.270$, $\Omega_\Lambda =0.73$, $h=0.72$.  Unless
otherwise specified, all distances are comoving\footnote{Note that in
  QPQ1 proper distances were quoted.}.  It is helpful to remember that
in the chosen cosmology, at a redshift of $z=2.5$, an angular
separation of $\Delta\theta=1\arcsec$ corresponds to a comoving
transverse separation of $R=20~\hkpc$, and a velocity difference of
$1500\kms$ corresponds to a radial redshift space distance of
$s=15~\hMpc$.  For a quasar at $z=2.5$, with an SDSS magnitude of
$r=19$, the flux of ionizing photons is 130 times higher than the
ambient extragalactic UV background at an angular separation of
$60\arcsec$ (comoving $R=1.2~\hMpc$).  Finally, we use the term optically thick
absorbers and LLSs interchangeably, both referring to quasar
absorption line systems with $\log N_{\rm HI}>17.2$, making them
optically thick at the Lyman limit ($\tau_{\rm LL}\gtrsim 1$).

\section{Quasar-Absorber Sample}
\label{sec:sample}

\begin{deluxetable*}{lcccccccccc}
\tablecolumns{11}
\tablewidth{0pc}
\tablecaption{Super-LLSs Near Quasars from QPQ1\label{table:slls}}
\tablehead{Name & z$_{\rm bg}$ & z$_{\rm fg}$ & $\Delta \theta$ & $R$ & z$_{\rm abs}$ & $\left|\Delta v\right|$ & $\Delta v_{\rm fg}$ & $\log N_{\rm HI}$ & $g_{\rm UV}$ & Telescope\\
                & &               &     ($^{\prime\prime}$)     &  ($\hkpc$) &      &  (km s$^{-1}$)           & (km s$^{-1}$)      &   (cm$^{-2}$)     &             &  } 
\startdata
SDSSJ0225$-$0739  & 2.99 & 2.440 &   214.0     &  4310     &2.4476   &  \phn690    &\phn500&  $19.55\pm 0.2$  & \phn\phn\phn5 &  SDSS\\ 
SDSSJ0239$-$0106  & 3.14 & 2.308 & \phn\phn3.7 & \phn\phn72&2.3025   &  \phn540    & 1500  &  $20.45\pm 0.2$  &  6369         &  Keck\\ 
SDSSJ0256$+$0039  & 3.55 & 3.387 &   179.0     &  4195     &3.387    &  \phn\phn20 & 1000  &  $19.25\pm 0.25$ &   \phn\phn20  &  SDSS\\ 
SDSSJ0338$-$0005  & 3.05 & 2.239 & \phn73.5    &  1415     &2.2290   &  \phn960    & 1500  &  $20.9 \pm 0.2$  &  \phn\phn 13  &  SDSS\\
\smallskip
SDSSJ0800$+$3542  & 2.07 & 1.983 & \phn23.1    &  \phn415  &1.9828   & \phn\phn40  &\phn300&  $19.0 \pm 0.15$ & \phn 488      &  Keck\\
SDSSJ0833$+$0813  & 3.33 & 2.516 &   103.4     &  2112     &2.505\phn&  \phn 980   &  1000 &  $19.45 \pm 0.3$ & \phn\phn 18   &  SDSS\\
SDSSJ0852$+$2637  & 3.32 & 3.203 &   170.9     &  3917     &3.211\phn&  \phn550    &  1500 &  $19.25 \pm 0.4$ & \phn\phn 13   &  SDSS\\
SDSSJ1134$+$3409  & 3.14 & 2.291 &   209.2     &  4073     &2.2879   &  \phn320    &   500 &  $19.5 \pm 0.3$  & \phn\phn 11   &  SDSS\\
SDSSJ1152$+$4517  & 2.38 & 2.312 &   113.4     &  2216     &2.3158   &  \phn370    &   500 &  $19.1 \pm 0.3$  & \phn\phn 30   &  SDSS\\
\smallskip
SDSSJ1204$+$0221  & 2.53 & 2.436 & \phn13.3    & \phn267   &2.4402   &  \phn370    &  1500 &  $19.7\pm 0.15$  & \phn 625      &  Gemini\\
SDSSJ1213$+$1207  & 3.48 & 3.411 &   137.8     &  3246     &3.4105   &  \phn\phn30 &  1500 &  $19.25\pm 0.3$  &  \phn\phn39   &  SDSS\\
SDSSJ1306$+$6158  & 2.17 & 2.111 & \phn16.3    & \phn302   &2.1084   &  \phn200    &   300 &  $20.3\pm 0.15$  & \phn 420      &  Keck\\
SDSSJ1312$+$0002  & 2.84 & 2.671 &   148.5     &  3129     &2.6688   &  \phn200    &   500 &  $20.3\pm 0.3$   & \phn\phn23    &  SDSS\\
SDSSJ1426$+$5002  & 2.32 & 2.239 &   235.6     &  4529     &2.2247   &   1330      &   500 &  $20.0\pm 0.15$  & \phn\phn19    &  SDSS\\
\smallskip
SDSSJ1430$-$0120  & 3.25 & 3.102 &   200.0     &  4517     &3.115\phn&  \phn960    &  1500 &  $20.5\pm 0.2$   &  \phn\phn 26  &  SDSS\\
SDSSJ1545$+$5112  & 2.45 & 2.240 & \phn97.6    &  1873     &2.243\phn&  \phn320    &   500 &  $19.45\pm 0.3$  &  \phn\phn30   &  SDSS\\
SDSSJ1635$+$3013  & 2.94 & 2.493 & \phn91.4    &  1861     &2.5025   &  \phn820    &   500 &  $>19$           &  \phn111      &  SDSS\\
\enddata
\tablecomments{\footnotesize
  Optically thick absorption line systems near foreground quasars. The
   background and foreground quasar redshifts are denoted by $z_{\rm
     bg}$ and $z_{\rm fg}$, respectively. The angular separation of the
  quasar pair sightlines is denoted by $\Delta\theta$, which
  corresponds to a transverse comoving separation of $R$ at the
  foreground quasar redshift.  Absorber redshift is indicated by
  $z_{\rm abs}$, and $\left|\Delta v\right|$ is the velocity difference between
  the absorber redshift and our best estimate of the redshift of the
  foreground quasar. Our estimated error on the foreground quasar
  redshift is denoted by $\Delta v_{\rm fg}$. Foreground quasar redshifts and 
  redshift errors were estimated according to the detailed procedure 
  described in \S~4 of QPQ1.  The logarithm of the column density of the
  absorber from a fit to the \ion{H}{1} profile is denoted by $\log
  \mnhi$. The column labeled ``Telescope'' indicates the instrument
  used to observe the background quasar.  The quantity 
  $g_{\rm UV}= 1 + F_{\rm QSO}\slash F_{\rm UVB}$ is the
  maximum enhancement of the quasars ionizing photon flux over that of
  the extragalactic ionizing background at the location of the background 
  quasar sightline, assuming that the quasar emission is isotropic 
  (see Appendix A of QPQ1). We compare to the UV background computed by F. 
  Haardt \& P. Madau (2006, in preparation)}
\end{deluxetable*}

In this section we briefly summarize the quasar-absorber sample and
the parent sample of projected pairs from which it was selected; for
full details see QPQ1. In \S~\ref{sec:cluster} we will quantify the
clustering of absorbers around quasars in the transverse direction by
comparing the number of quasar-absorber pairs discovered to the number
from random expectation. Although lower column density systems were
published in QPQ1, we focus here on absorbers with $\mnhi >
10^{19}{\rm cm}^{-2}$, or so called super-LLSs. This choice reflects a
variety of concerns. First, we wanted to maximize the number of
quasar-absorber pairs -- had we restricted consideration to only DLAs
we would have been left with 5 systems. Second, this is the lowest
column density at which we believe we can identify a statistical
sample of absorbers with the moderate resolution spectra used in
QPQ1. Finally, the column density distribution of optically thick
absorption line systems has been determined for $\log \mnhi > 19$
\citep{Peroux05,Omeara06}; whereas, it is unknown at the lower column
densities $17.2 \lesssim \log \mnhi \lesssim 19$.

Modern spectroscopic surveys select \emph{against} close pairs of
quasars because of fiber collisions.  For the Sloan Digital Sky Survey
(SDSS), the finite size of optical fibers implies only one quasar in a
pair with separation $<55\arcsec$ can be observed spectroscopically on
a given plate\footnote{An exception to this rule exists for a fraction
  ($\sim 30\%$) of the area of the SDSS spectroscopic survey covered
  by overlapping plates.  Because the same area of sky was observed
  spectroscopically on more than one occasion, there is no fiber
  collision limitation.}, and a slightly smaller limit ($<30\arcsec$)
applies for the Two Degree Field Quasar Survey (2QZ) \citep{Croom04}.
Thus, for sub-arcminute separations, additional spectroscopy is
required both to discover companions around quasars and to obtain
spectra of sufficient quality to search for absorption line systems.
For wider separations, projected quasar pairs can be found directly in
the SDSS spectroscopic quasar catalog.  \citet{binary} used the 3.5m 
telescope at Apache Point Observatory (APO) to spectroscopically confirm a 
large sample of photometrically selected close quasar pair candidates, and 
published the largest sample of projected quasar pairs in existence.  

In QPQ1 we combined high signal-to-noise ratio (SNR) moderate
resolution spectra of the closest \citet{binary} projected pairs,
obtained from Gemini, Keck, and the Multiple Mirror Telescope (MMT),
with a large sample of wider separation pairs, from the SDSS
spectroscopic survey.  The Keck, Gemini, and MMT spectra were all of
sufficient signal-to-noise ratio to detect absorbers with column
densities $\mnhi > 10^{19}{\rm cm}^{-2}$, and a SNR criterion was used
to isolate the SDSS projected pair sightlines for which such an
absorber could be detected in the background quasar spectrum. We
considered all projected quasar pair sightlines which had a comoving
transverse separation of $R < 5~\hMpc$, a redshift separation $\Delta
v > 2500\kms$ between the foreground and background quasar (to exclude
physical binaries), and which satisfied our SNR criteria, and arrived
at a total of 149 projected pair sightlines in the redshift range $1.8
< z < 4.0$.

%Keck spectra accounted for 25 of the background quasar
%spectra, 19 came from Gemini, 2 from the MMT and the remaining 103
%were from the SDSS.  Of these sightlines, 25 had angular separations
%below the SDSS fiber collision limit ($\Delta \theta < 55\arcsec$)
%with the Keck/Gemini/MMT accounting for all but three of these.

A systematic search for optically thick absorbers with redshifts
within $|\Delta v| < 1500\kms$ of the foreground quasars was conducted
by visually inspecting the 149 projected pair sightlines, where the
velocity window was chosen to bracket the uncertainties in the
foreground quasars' systemic redshift \citep[see e.g.][and \S~4 of
  QPQ1]{Richards02}. Voigt profiles were fit to systems with
significant Ly$\alpha$ absorption to determine the \ion{H}{1} column
densities.  We uncovered 27 new quasar absorber pairs with column
densities $17.2 < \log \mnhi < 20.9$ and transverse comoving distances
$71~\hkpc < R < 5~\hMpc$ from the foreground quasars, of which 17 were
super-LLSs with $\log \mnhi > 19$.  The mean redshift of the
foreground quasar in the parent sample was $\langle z\rangle = 2.47$
and that of the 17 quasar-super-LLS pairs was $\langle z_{\rm
  abs}\rangle = 2.55$.

The completeness and false positive rate of the QPQ1 survey are a
significant source of concern. \cite{phw05} demonstrated that the
spectral resolution (FWHM $\simeq 150\kms$) and SNR of the SDSS
spectra are well suited to constructing a complete ($\gtrsim 95\%$)
sample of DLAs ($\log \mnhi > 20.3$) at $z > 2.2$; however, the
completeness at the lower column densities $\log \mnhi > 19$,
considered here has not been systematically quantified.  Furthermore,
aggressive SNR criteria were employed in QPQ1 in order to gather a
sufficient number of projected quasar pairs. Line-blending can
significantly depress the continuum near the \lya\ profile and mimic a
damping wing, biasing column density measurements high or giving rise
to false positives. Any statistical study will thus suffer from a
`Malmquist'-type bias because line-blending biases lower column
densities upward, and the line density of absorbers $dN\slash dz$, is
a steep function of column density limit.

Based on visually inspecting the 149 background quasar spectra and a
comparison with echelle data for three systems, we estimated that the
QPQ1 survey was $\sim 90\%$ complete for $\log \mnhi > 19.3$ for all
the Keck/Gemini/MMT spectra and $\sim 3/4$ of the SDSS spectra, which
accounts for about 125 of the 149 spectra searched.  To address the
false positive rate, we compared with echelle data for three systems
and found that our column density was overestimated by $\sim
2.5\sigma$ for one system, raising it above the super-LLS ($\log \mnhi
> 19$) threshold. However, this absorber was located blueward of the
quasars Ly$\beta$\ emission line, in a part of the spectrum `crowded'
by the presence of both the Ly$\alpha$\ and Ly$\beta$ forests. A more
careful examination of the completeness and false positive rate of
super-LLSs identified in spectra of the resolution and SNR used in
QPQ1 is definitely warranted.  In \S~\ref{sec:cluster} we explore how
our clustering measurement changes if we discard systems within
$\approx 1 \sigma$ of the $\log \mnhi = 19$ threshold.

%A discussion of how the foregroud quasar systemic redshifts were
%estimated is given in \S~4 of QPQ1. We briefly summarize the essential
%details. The emission line centers of \ion{C}{4}, \ion{C}{3}, and
%\ion{Mg}{2} were measured if they landed within the wavelength
%coverage of the foreground quasar spectrum.  Emission lines were
%centered using the relation mode = 3 $\times$ median - 2$\times$ mean,
%which is a more robust estimator than the centroid or median for
%slightly skewed profiles in noisy data.  Redshifts were assigned using
%the emission lines which tend to show the least dispersion and
%smallest shifts about systemic \citep[see e.g. ][]{Richards02} Thus
%the \ion{Mg}{2} redshift was used if the emission line was covered by
%the spectrum with sufficient SNR. Otherwise, \ion{C}{3} was used or
%\ion{C}{4}, if neither \ion{Mg}{2} or \ion{C}{3} were
%present. Redshift errors were assigned based on the measured
%dispersion of emission line centeres about systemic measured by
%\citet{Richards02}.  In some cases where none of the aformentioned
%lines were covered by the spectrum, the redshifts were mesaured by
%cross-correlating the spectrum with a template quasar spectrum.

Relevant quantities for the super-LLS-quasar pairs which are used in
our clustering analysis are given in Table~\ref{table:slls}.  The
distribution of foreground quasar redshifts, transverse separations,
and ionizing flux ratios probed by all of our projected pair sightlines is
illustrated in Figure~\ref{fig:scatter}.  The filled symbols outlined
in black indicate the sightlines which have an absorption line system
near the foreground quasar with $\log \mnhi > 19$ (see
Table~\ref{table:slls}) and open symbols are sightlines with no such 
absorber.

\section{Quantifying Quasar-Absorber Clustering}
\label{sec:formalism}

%\begin{figure}
%  \centerline{\epsfig{file=chi_trans.ps,bb=70 70 560 420
%      ,width=0.50\textwidth}}
%  \caption{Transverse correlation function \label{fig:trans}}
%\end{figure}

\begin{figure}
  \centerline{\epsfig{file=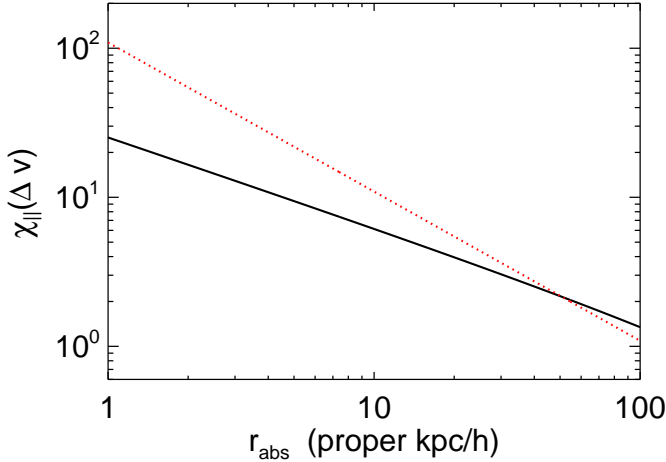,bb=110 70 570 420
      ,width=0.50\textwidth}}
  \caption{ Dependence of line-of-sight correlation function on the
    (proper) size of the absorber cross section $r_{\rm abs}$.  A
    velocity interval of $\Delta v = 3000~\kms$ was assumed. Small
    cross-sections sample smaller scales near the quasar where the
    correlation $\xi_{\rm QA}\propto r^{-\gamma}$ function is large,
    giving a larger value for the average in eqn.~(\ref{eqn:pdla}) for
    $\chi_{\parallel}$; whereas, large cross-sections dilute $\xi_{\rm
      QA}$ in the average, giving smaller values of
    $\chi_{\parallel}$. The solid (black) line is for a correlation
    length of $r=4.7~\hMpc$ and slope of $\gamma=1.6$.  The dotted
    (red) line is for a correlation function with slope $\gamma=2.0$,
    normalized to the have the same clustering amplitude as the
    the shallower $\gamma=1.6$ model at $r = 1~\hMpc$. \\
    \label{fig:pdla}}
\end{figure}

In the absence of clustering, the line density of absorption line
systems per unit redshift above the column density threshold $N_{\rm
  HI}$ is given by the cosmic average 
\be
\left\langle\frac{dN}{dz}\right\rangle(> N_{\rm HI},z) = n A f_{\rm
  cov} \frac{c}{H(z)}\label{eqn:dndz}, 
\ee 
where $n$ is comoving
number density of the galaxies or objects which give rise to
absorption line systems, $A$ is their absorption cross section (in
comoving units), $f_{\rm cov}$ is the covering factor, and $H(z)$ is
the Hubble constant.  Note that the line density is degenerate with
respect to the combination $nAf_{\rm cov}$ and only their product can
be determined by measuring the abundance of absorption line systems.

At an average location in the Universe, the probability of finding a
absorber in a background quasar spectrum within the redshift
interval $\Delta z = 2(1+z)\Delta v\slash c$, corresponding to a
velocity interval $2\Delta v$ is simply $P=\langle dN\slash dz \rangle
\Delta z$.  For a projected pair of quasars, clustering around the foreground
quasar will increase the probability of finding an absorber in the 
background quasar spectrum in the vicinity of the foreground quasar; 
whereas, the foreground quasars radiation field could reduce this enhancement
by photoevaporating absorption systems. If the quasar sightlines have a 
comoving transverse separation  $R$, and 
assuming that one searches a velocity interval $\pm \Delta v$
about the foreground quasar redshift (because of redshift
uncertainties), we can express the increase in line density near the
foreground grounds in terms of a transverse correlation function
$\chi_{\rm \perp}(R)$ as
\be 
\frac{dN}{dz}=
\left\langle\frac{dN}{dz}\right\rangle \left[1 + \chi_{\perp}(R,\Delta v)\right]
\label{eqn:clust}
\ee 
where $\chi_{\rm \perp}(R,\Delta v)$ is given by an average of the 3-d
quasar-absorber correlation function, $\xi_{\rm QA}(r)$, over a
cylinder with volume $V=A[2\Delta v\slash a H(z)]$. Here, $a$
is the scale factor and $2\Delta v (1+z)\slash H(z)$ is the length of
the cylinder in the LOS direction. We can thus write
\bea
\chi_{\perp}(R) & = & \frac{1}{V}\int_{\rm
  V}dV~\xi_{\rm QA}(r)\\
& \approx & \frac{a H(z)}{2\Delta v}
\int_{\frac{-\Delta v}{a H(z)}}^{\frac{\Delta v}{a H(z)}} 
dZ~\xi_{\rm QA}(\sqrt{R^2+ Z^2})\nonumber,  
\label{eqn:trans} 
\eea
where $Z$ is a comoving distance in redshift space.  The last
approximation in eqn.~\ref{eqn:trans}
assumes that the volume average over the cylinder can be
replaced by the line average along the line-of-sight direction, which
is valid provided that we are in the `far-field' limit, i.e., the
transverse separation is much larger than the diameter of the cylinder
$R \gg {\sqrt A}$. Thus provided we consider distances $R$ much larger
than the dimension of the absorber, the transverse clustering is
independent of the absorption cross-section.

For proximate absorbers along the line of sight, we can similarly
define a line-of-sight correlation function 
\be
dN\slash dz = \langle dN\slash dz \rangle[1+ \chi_{\parallel}(\Delta v)]
\label{eqn:pdla_dndz}
\ee
where
\be
\chi_{ \parallel}(\Delta v)  =  \frac{aH(z)}{A\Delta v}
\int_{Z_{\rm cut}}^{\frac{\Delta v}{aH(z)}} dZ \int dA
~\xi_{\rm QA}(r). 
\label{eqn:pdla}
\ee
The lower limit of the radial integration is set to $Z_{\rm cut}$,
a cutoff which is introduced to parametrize our ignorance of the
geometry of the absorbers. For instance, if the absorbers were pancake 
shapes which were always oriented perpendicular to the line of sight 
(like face-on spiral galaxies) then we would not cutoff the line-of-sight 
integration at all ($Z_{\rm cut}=0$), since face on pancakes at zero 
separation can still obscure the quasar. For 'hard sphere', we would 
set $Z_{\rm cut}=2 \sqrt{A\slash \pi}$ --- presuming that the obscuring 
cross-section of the absorber drops to zero for points interior to it. 
As smaller values of $Z_{\rm cut}$ will correspond to larger
line-of-sight clustering, we henceforth conservatively assume the hard sphere
case and use $Z_{\rm cut}=2\sqrt{A\slash\pi}$

Note that one is no longer in the `far-field' limit for the integral
in eqn.~(\ref{eqn:pdla}), and the clustering amplitude
$\chi_{\parallel}$ explicitly depends on the cross section of the
absorber. It is easy to understand the nature of this dependence if 
one considers that the line density $\propto n A$ is fixed by
measurements of $\langle dN\slash dz\rangle$. Thus, smaller cross-sections
correspond to larger volume number densities, but smaller scales near
the quasar are being sampled by the integral in eqn.~(\ref{eqn:pdla}), where
the correlation $\xi_{\rm QA}\propto r^{-\gamma}$ function is large.
Conversely, larger cross-sections sample regions further from the
quasar, and thus $\xi_{\rm QA}$ is averaged over a larger volume and
hence diluted.  Figure~\ref{fig:pdla} illustrates the dependence of
$\chi_{\parallel}(\Delta v)$ on the proper radius of the absorber cross 
section, $r_{\rm abs}$, for a range of sizes.

In eqns.~(\ref{eqn:trans}) and (\ref{eqn:pdla}) $\xi_{\rm QA}(r)$
represents a real space correlation function and it may appear that we
have neglected redshift space distortions.  Strictly speaking, the
redshift space correlation function $\xi_{\rm QA}(R,Z)$, should be
appear in eqns.~\ref{eqn:trans} and \ref{eqn:pdla}. This $\xi_{\rm
  QA}(R,Z)$ is the convolution of the real space correlation function
$\xi(r)$, with the velocity distribution in the radial direction,
which can have contributions from both peculiar velocities and
uncertainties in the systemic redshifts of the quasar.  However,
provided that the distance in redshift space over which we project
$\Delta v$ contains most of the probability under this distribution, it is
a good approximation to replace the redshift space correlation
function, under the integrals in eqns.~(\ref{eqn:trans}) and
(\ref{eqn:pdla}), with the real space correlation function, because radial
velocities will simply move pairs of points within the volume.

% On small scales of interest to us here, 
% neglecting redshift space distortions would lead us to 
% \emph{underestimate} the clustering, because infall and virialization will
%give rise to a `finger of god' enhancement along the line of sight. 

% The associated unsaturated metal lines in high column density systems show
% multiple components or clouds. This kinematic structure is casued by
% peculiar velocities, not Hubble flow. They are thought to represenet
% turbulent or thermal motions in the gas (Prochaska \& Wolfe
% 199?). These velocities limit the precision with which the systemic
% redshifts of absorption line systems can be determined, however this
% precision will depend on the kinematic structure of the metal
% lines. If the metals indicate a single cold absorber, as is the case
% in some DLAs, then the redshift could be determined to better than $20
% \kms$. However, the situation could be as bad as multiple absorbers
% spread by $\sim 500\kms$ (refs), in which case the systemic redshift
% cannot be determined to much better than $\sim 500\kms$.
% Uncertainties in the quasar redshift can be as large as $\sim
% 2000\kms$, if the redshift is determined from high ionization broad
% emission lines, which tend to be significantly blue-shifted from the
% systemic redshift. This uncertaintly can be mitigated by using lower
% ionization lines (like MgII, OII OIII), but for high redshift quasars 
% $z\gtrsim 2.5$ this requires near-infrared spectroscopy. 

% LEFT OFF HERE IN REREAD

\section{Estimating the Correlation Function}
\label{sec:maxL}
%In this section we introduce a simple maximum likelihood estimator for
%the transverse quasar absorber correlation function and we describe monte
%carlo simulations which show that this estimator is unbiased. These monte
%carlo simulations are used to asssign errors to our clustering measurement
%in~\S~\ref{sec:cluster}.  We also described an estimator for the correlation
% function in transverse distance bins. In order to compute the random number 
% of absorbers expected 

\subsection{Maximum Likelihood Estimator}

Given a quasar-absorber correlation function $\xi_{\rm QA}$,
eqns.~(\ref{eqn:clust}) and (\ref{eqn:trans}) describe how to compute the
probability of finding an absorber in the redshift interval $\Delta z
= 2(1+z)\Delta v\slash c$ at a transverse distance $R$ from a
foreground quasar: $P(R,z)= (dN\slash dz) \Delta z$.
Considering that we only have 17 quasar-absorber pairs selected from 
149 sightlines, it will be difficult to measure more than a single 
parameter with reasonable errors. Hence, we assume the quasar-absorber 
correlation function to have a power-law form 
\be
\xi_{\rm QA} = C\left(\frac{r}{r_0}\right)^{-\gamma}, 
\ee 
where $C$ is a clustering amplitude and $r_0$ is the correlation
length.  The amplitude $C$ is degenerate with $r_0$, but we choose to
estimate $C$, because it allows for the possibility of
anti-correlation ($C < 0$) which could result from the QSO ionizing
radiation field. Motivated by the the slope of the LBG
auto-correlation function \citep{Adel05a}, we choose to fix
$\gamma=1.6$.  A similar procedure was employed by \citep[][
henceforth AS05]{AS05}, who measured clustering of LBGs around
luminous quasars ($2 \lesssim z \lesssim 3.5$), and found a best fit
correlation length of $r_0=4.7~\hMpc$. We set $r_0=4.7~\hMpc$ as a
fiducial value; thus $C$ can be interpreted as the quasar-absorber
clustering amplitude relative to the AS05 quasar-LBG result.

Consider an ensemble of $N$ projected pair sightlines with background
quasars at transverse separations $R_i$ from foreground quasars at
redshifts $z_i$. Given $\xi_{QA}$, we can compute the 
associated probabilities $P_i \equiv P(R_i,z_i)$.
Suppose that $N_{\rm SLLS}$ of these $N$ sightlines show absorption
from super-LLSs with $\log \mnhi > 19$. The likelihood of the data,
given the model parameter $C$ is then 
\be
\mathcal{L}(C)=\prod_{i}^{N_{\rm SLLS}} P_i \prod_j^{N-N_{\rm SLLS}}
(1-P_j)\label{eqn:ML}, 
\ee 
where the probabilities $P_i\le 1$ are capped at 1. By maximizing the 
likelihood with respect to the parameter $C$, we estimate the 
clustering amplitude from the data.

\subsection{Monte-Carlo Simulations}
\label{sec:monte}
We must verify that the maximum likelihood estimator in
eqn.~(\ref{eqn:ML}) is unbiased, and we would also like to know how 
to assign error bars to an estimate of $C$. Both of these 
points can be addressed with monte-carlo methods.  The distribution of 
redshifts and transverse separations in Figure~\ref{fig:scatter} is not 
uniform, and it is thus important that we preserve this distribution when 
constructing mock data sets to assign errors. 

For a `true' value of the clustering amplitude $C$, we can compute the
probabilities $P_i \equiv P(R_i,z_i)$ of observing an absorber for each
of the $N$ projected pair sightlines. Mock data sets can then be
constructed by generating an $N$-dimensional vector of deviates from
the uniform distribution $x_i$, and assigning sightlines with $x_i <
P_i$ an absorption line system.  The same maximum likelihood estimator
is applied to these mock data sets to determine an estimate ${\hat
  C}$ for each one, thus allowing us to measure the probability
distribution $P({\hat C}|C)$ of the estimate about the true model.

\subsection{Binned Correlation Function}

An alternative to the maximum likelihood technique described above, would
be to estimate the correlation length by fitting a power-law model to
estimates of the correlation function in bins of transverse
separation. However, because we have only 17 quasar-absorber pairs,
the best-fit correlation length from this procedure would be very
sensitive to the binning chosen.  We compute the correlation function
in bins because it provides a useful way to visualize the data,
although we do not fit the binned data for $C$

A simple estimator for the transverse correlation function in 
eqn.~(\ref{eqn:clust}) is \footnote{This is actually an estimator of
  $\chi_{\perp}(R,\Delta v)$ averaged over the volume of the bin. We
  ignore this distinction, because the differences are substantially
  smaller than our errors and we are not fitting parameters from this
  estimate.}
\be
\chi_{\perp}(R,\Delta v) = \frac{\langle {\rm QA}\rangle}{\langle {\rm QR}\rangle} - 1
\label{eqn:chibin}, 
\ee 
where $\langle {\rm QA}\rangle$ is the number of quasar-absorber
pairs in a transverse radial bin centered on $R$, and $\langle {\rm
  QR}\rangle = \sum_i \langle dN\slash dz\rangle \Delta z$, is the number
of quasar-random pairs expected. 

\subsection{The Column Density Distribution}
\label{sec:omeara}

Before we can estimate the correlation amplitude with
eqn.~(\ref{eqn:ML}), we require the cosmic average line density of
super-LLSs $\langle dN\slash dz\rangle$. The line density is the zeroth
moment of the column density distribution 
\be 
\left\langle \frac{dN}{dz}\right\rangle (> \mnhi,z)  = \int_{\mnhi}^{\infty} 
f_{\rm HI}(\mnhi,z)d\mnhi \label{eqn:fofn}.
\ee
\citet{Omeara06} measured the column density distribution for
super-LLSs in the range $19 < \log \mnhi < 20.3$ at $z \sim 2.7$ and
found good agreement with a power law $f(N) = A (\mnhi\slash
10^{19}~{\rm cm^{-2}})^b$, where $A=1.10\times 10^{-20}$ and
$b=-1.43$. For $\log \mnhi > 20.3$, \citep{phw05} measured the column 
density distribution from the SDSS quasar sample in redshift bins. 

To evaluate the integral in eqn.~(\ref{eqn:fofn}), we use the
\citet{Omeara06} power law fit in the range $19 < \log \mnhi < 20.3$,
and a spline fit to the \citet{phw05} results in the redshift bin
centered at $z\sim 2.7$. For the redshift evolution, we simply scale
the line density of all absorbers by the evolution of DLAs 
$\log \mnhi >20.3$ measured by \citet{phw05} (see their Figure~8). This
procedure assumes that the abundance of super-LLSs evolves similarly to 
that of DLAs, an assumption that is consistent with but not 
confirmed by \citet{Omeara06}.

\section{Clustering Results}
\label{sec:cluster}

\begin{figure}
  \centerline{\epsfig{file=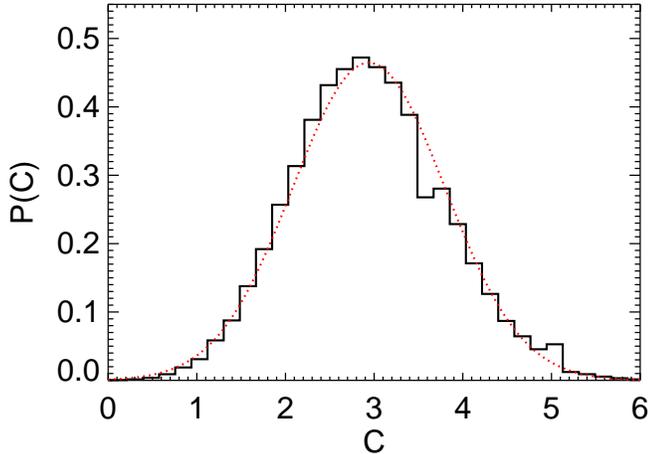,bb= 40 0 500 350,width=0.50
      \textwidth}}
  \caption{Distribution of clustering amplitude estimates from monte-carlo
    simulations. The black histogram shows the probability distribution of the
    clustering amplitude maximum likelihood estimates $P({\hat C})$, for a
    model with true value equal to our measurement of $C=2.9$. This
    distribution was created by applying the maximum likelihood estimator in 
    eqn.~(\ref{eqn:ML}) to 100,000 mock realizations of our 149 projected pair 
    sightlines, as described in \S~\ref{sec:monte}. The mean of this 
    distribution is $\langle {\hat C}\rangle = 2.94$, the dispersion is 
    $\sigma = 0.86$, and the (red) dotted curve shows a Gaussian with the 
    same mean and dispersion. The full distribution (black histogram) is
    used to assign errors to our measurement.\label{fig:dist}}
\end{figure}

\begin{figure}
  \centerline{\epsfig{file=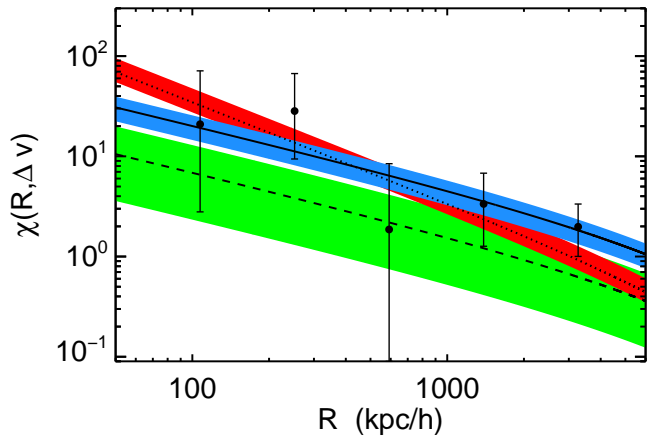,bb=80 90 570 400,
      width=0.5\textwidth}}
  \caption{Binned transverse quasar-absorber correlation function for
    17 quasar-absorber pairs selected from 149 projected pair
    sightlines.  The lines indicate the best-fit model from our
    maximum likelihood analysis and the shaded regions indicate the
    range allowed by $1\sigma$ errors estimated from monte-carlo
    simulations.  The solid line and the blue shaded region are for an
    assumed correlation function slope of $\gamma=1.6$. The steeper
    dotted line and red shaded region are for $\gamma=2.0$. The dashed
    curve and green region show the transverse correlation function
    $\chi_{\perp}$ if we set the quasar-absorber correlation
    function $\xi_{\rm QA}$ to the QSO-LBG correlation
    function measured by AS05 ($r_0 = 4.7\pm 1.3~\hMpc;
    \gamma=1.6$)\label{fig:corr}}
\end{figure}

We applied the maximum likelihood estimator to the 17 quasar-absorber pairs 
(see Table~\ref{table:slls}) selected from the 149 projected pair sightlines
shown in Figure~\ref{fig:scatter}. The maximum likelihood value of the
clustering amplitude is $C=2.9$, which corresponds to a correlation
length of $r_0 = 9.2~\hMpc$.

In Figure~\ref{fig:dist} we show the probability distribution of the
clustering amplitude maximum likelihood estimates $P({\hat C})$, for a
model with true value equal to our measurement of $C=2.9$. This
distribution was created by applying the maximum likelihood estimator in 
eqn.~(\ref{eqn:ML}) to 100,000 mock realizations of our 149 projected pair 
sightlines, as described in \S~\ref{sec:monte}. The mean of this distribution is
$\langle {\hat C}\rangle = 2.94$, the dispersion is $\sigma = 0.86$,
and the (red) dotted curve shows a Gaussian with the same mean and
dispersion. The monte-carlo simulation indicates that our estimator is
indeed unbiased and that the error distribution is reasonably well
approximated by a Gaussian. Using the full distribution from the monte-carlo
simulation, we compute the $68\%$ confidence interval about the `true' 
input value, which we use to quote errors on our clustering measurement: $C
= 2.9\pm 0.8$, or $r_0=9.2^{+1.5}_{-1.7}~\hMpc$.  If we assume a steeper 
slope for the correlation function $\gamma=2$,  we obtain 
$r_0=5.8^{+1.0}_{-0.6}~\hMpc$.

% <<<<<<< clust.tex
% approximated by a gaussian. We will thus use the standard deviation
% from the monte-carlo to quote errors on our clustering measurement: $C
% = 2.93\pm 0.86$, or $r_0=9.2^{+1.6}_{-1.8}\hMpc$.

% We detect strong clustering of super-LLSs around quasars, assuming
% $\gamma = 1.6$, our best fit clustering amplitude is three times
% larger than the clustering of LBGs around quasars measured by
% \citet{AS05}. The 4/8 small scale ($R < 500\hkpc$) absorbers could
% be driving this large amplitude. If we omit the 8 sightlines with 
% $R < 500\hkpc$, and redo the analysis we measure the smaller value 
% $r_0=7.6^{+1.9}_{-2.2}\hMpc$. 

% If we assume we overestimated all the colum densities by $0.3$~dex
% because of line blending, such that our effective limiting 
% column density is $\log \mnhi = 18.7$ instead of $\log \mnhi = 19$, we
% obtain $r_0=7.1^{+}_{-}\hMpc$. 

% If we assume a  steeper slope $\gamma=2$ we measure 
% $r_0=5.8^{+0.85}_{-1.0}\hMpc$. 

% Plot binned correlation function and max-L fits. Also try fitting a 
% $X(R,\Delta z)$ parameterized function to quote? Choose a $\Delta z$ and 
% a power law slope and fit for the amplitdue? 

% =======

We detect strong clustering of super-LLSs around quasars.
Figure~\ref{fig:clust} shows the binned transverse correlation
function $\chi_{\perp}$ compared to the two maximum likelihood fits
($\gamma=1.6$ and $\gamma=2$) and to the prediction from the
quasar-LBG correlation function measured by AS05.  For $\gamma = 1.6$,
the quasar-absorber clustering amplitude is three times larger than
the quasar-LBG measurement of AS05 and inconsistent with it at the
$\gtrsim 1\sigma$ level. 

The auto-correlation functions of high redshift galaxies tend to
become progressively steeper on (comoving) scales $R\lesssim 1~\hMpc$
characteristic of the sizes of dark matter halos 
\citep{Coil05,Ouchi05,Lee06,Conroy06}, thus similar behavior might be in
cross-correlation around quasars. Indeed, \citet{binary} detected an
order of magnitude excess quasar auto-clustering on scales $\lesssim
100~\hkpc$. For this reason, we also quote our results for $\gamma=2$,
although a steeper slope is not clearly favored by the binned data in
Figure~\ref{fig:corr}. 

At first glance, it may seem odd that the relative error on our
measurement of $r_0$ is $\sim 20\%$ using only 17 quasar-absorber
pairs; whereas AS05 quote a $\sim 30\%$ error on $r_0$ from $\sim 200$
quasar-LBG pairs around a sample of $\sim 50$ quasars. The estimator
used here is qualitatively similar to that used by AS05 \citep[see
also][]{Adel05a} in that both techniques rely on the radial clustering
of pairs of objects at fixed angular positions, because the angular
selection functions are unknown \citep[see][]{Adel05b}. It is thus worth
explaining that our smaller errors arise from a few factors.

%A comparison
%of relative errors in the cross-correlation of different objects
%(quasar-LBG versus quasar-absorber) determined from different
%techniques and different samples may seem to be of limited
%use. 

First, the relative error decreases with clustering amplitude, and our
best fit has a factor of three larger clustering strength. Second,
AS05 exclude scales $R\lesssim 1.2~\hMpc$ from their analysis; whereas
four out of our 17 quasar-absorber pairs have $R<1.2~\hMpc$. Small
scale pairs are effectively `worth' many large scale pairs because the
signal to noise per pair is much higher. Finally, the AS05 errors
are determined from the field to field dispersion in the data, which 
includes a contribution from cosmic variance. Because the projected pairs
of quasars are distributed over the entire sky, our measurement does
not suffer from cosmic variance errors. 

%If we exclude pair
%sightlines with $R < 1.2~\hMpc$ and redo our analysis we measure
%weaker clustering $r_0=6.9^{+1.9}_{-2.2}\hMpc$, with a larger ($\sim
%30\%$) relative error.

To illustrate that the errors from our monte-carlo technique
are sensible and comparable to the AS05 result, we randomly resampled
our pair sightlines with $R > 1.2~\hMpc$ to create a mock projected
pair sample 30 times larger, but with the same distribution of
redshifts and transverse distances. We increased the search window
$\Delta v=3000~\kms$, in closer agreement with the $l=30~\hMpc$ radial
window averaged over by AS05.  The average number of quasar-absorber
pairs expected from this hypothetical enlarged sample is $\langle N
\rangle = \sum_i P_i = 206$, which is of order the number of LBG-AGN
pairs used by AS05. Assuming $\gamma=1.6$ and $r_0=4.7~\hMpc$, our
monte-carlo simulation gives $r_0=4.7^{+0.9}_{-1.0}~\hMpc$, or a relative
error of $\sim 20\%$, comparable to the AS05 errors but slightly smaller. 

% Third, the
%estimator used to measure the quasar-LBG clustering, does not make
%full use of angular information, because the angular selection
%function of their survey is unknown \citep[see ][for
%  details]{Adel05}. 
%The 4/8 small scale ($R < 500\hkpc$) absorbers could be driving this
%large amplitude. If we omit the 8 sightlines with $R < 500\hkpc$, and
%redo the analysis we measure the smaller value
%$r_0=7.6^{+1.9}_{-2.2}\hMpc$.

\subsection{Systematic Errors}

\subsubsection{`Malmquist' Bias}

As discussed in \S~\ref{sec:sample}, the identification of $\log \mnhi
= 19$ in spectra at the SNR and resolution used in QPQ1, can result in
a significant `Malmquist' type bias because line-blending scatters
lower column density absorbers upward, and the line density of
absorbers $dN\slash dz$, is a steep function of column density
limit. If absorbers with column densities $\log \mnhi < 19$ scatter up
into our sample, this error would bias our clustering measurement
high. To investigate the impact of this bias, we redo the clustering
analysis but ignoring the six quasar-absorber pairs in
Table~\ref{table:slls} which have column densities within $1\sigma$ of
the threshold $log \mnhi = 19$: SDSSJ~0256$+$0039, SDSSJ~0800$+$3542,
SDSSJ~0852$+$2637', SDSSJ~1152$+$4517, SDSSJ~1213$+$1207, and
SDSSJ~1635$+$3013. The eleven remaining quasar-absorber pairs give a
maximum likelihood clustering amplitude of $C=1.7\pm 0.7$ or
$r_0=6.4^{+1.7}_{-1.8}~\hMpc$, compared to our measurement of
$C=2.9\pm 0.8$, or $r_0=9.2^{+1.5}_{-1.7}~\hMpc$ for the full
sample. This illustrates that even under very conservative assumptions
about the possible effect of Malmquist bias, the clustering amplitude
would be only be reduced by $\sim 1.5\sigma$.

\subsubsection{Redshift Errors}

Another possible source of bias in our sample could arise from the
determination of the foreground quasar redshifts.  Note that the
clustering analysis properly takes the redshift uncertainties into
account, by averaging the correlation function over a window of
$|\Delta v|=1500~\kms$. However, assigning redshifts to the foreground
quasar (see \S~4 of QPQ1) is not a completely objective process. The
quasar emission lines sometimes exhibit mild-BAL or metal line
absorption, and the line centering can be sensitive to how these
features are masked.  It is possible that we were biased towards
including quasar-absorber pairs in our sample, and thus tended to
assign redshifts resulting in velocity differences $|\Delta
v|<1500~\kms$.  To address this issue we redo the analysis discarding
the three quasar-absorber pairs in Table~\ref{table:slls} which have
velocity differences larger than the quoted error, $\Delta v_{\rm fg}
> |\Delta v|$: SDSSJ~0225$-$0739, SDSSJ~1426$+$5002, and
SDSSJ~1635$+$3013.  The 14 remaining quasar-absorber pairs give a
maximum likelihood clustering amplitude of $C=2.4^{+0.8}_{-0.7}$ or
$r_0=8.1^{+1.6}_{1.7}~\hMpc$. Furthermore, if we discard the five
quasar-absorber pairs with the largest velocity differences, we
measure $C=1.9\pm 0.7$ or $r_0=7.1^{+1.6}_{-1.7}~\hMpc$.  Thus we conclude that
even if we were biased towards including several quasar-absorber pairs
with the velocity differences $|\Delta v| > 1500~\kms$, this would
only reduce the the measured clustering amplitude by $\sim 1\sigma$.

\section{Anisotropic Clustering of Absorbers Around Quasars}
\label{sec:proximity}

\begin{figure*}
  \centerline{
    \epsfig{file=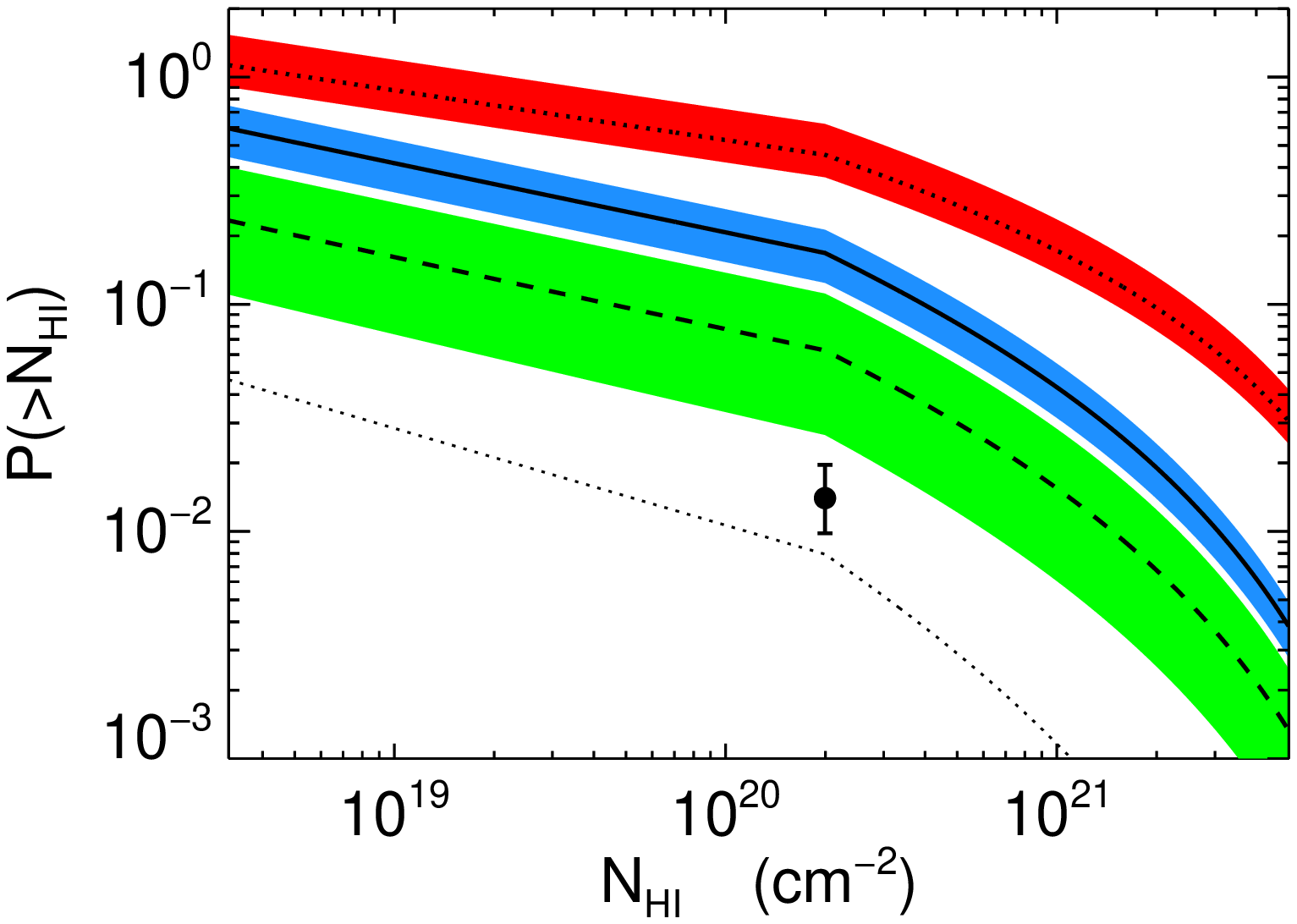,bb=82 72 576 412,
      width=0.5\textwidth}    
    \epsfig{file=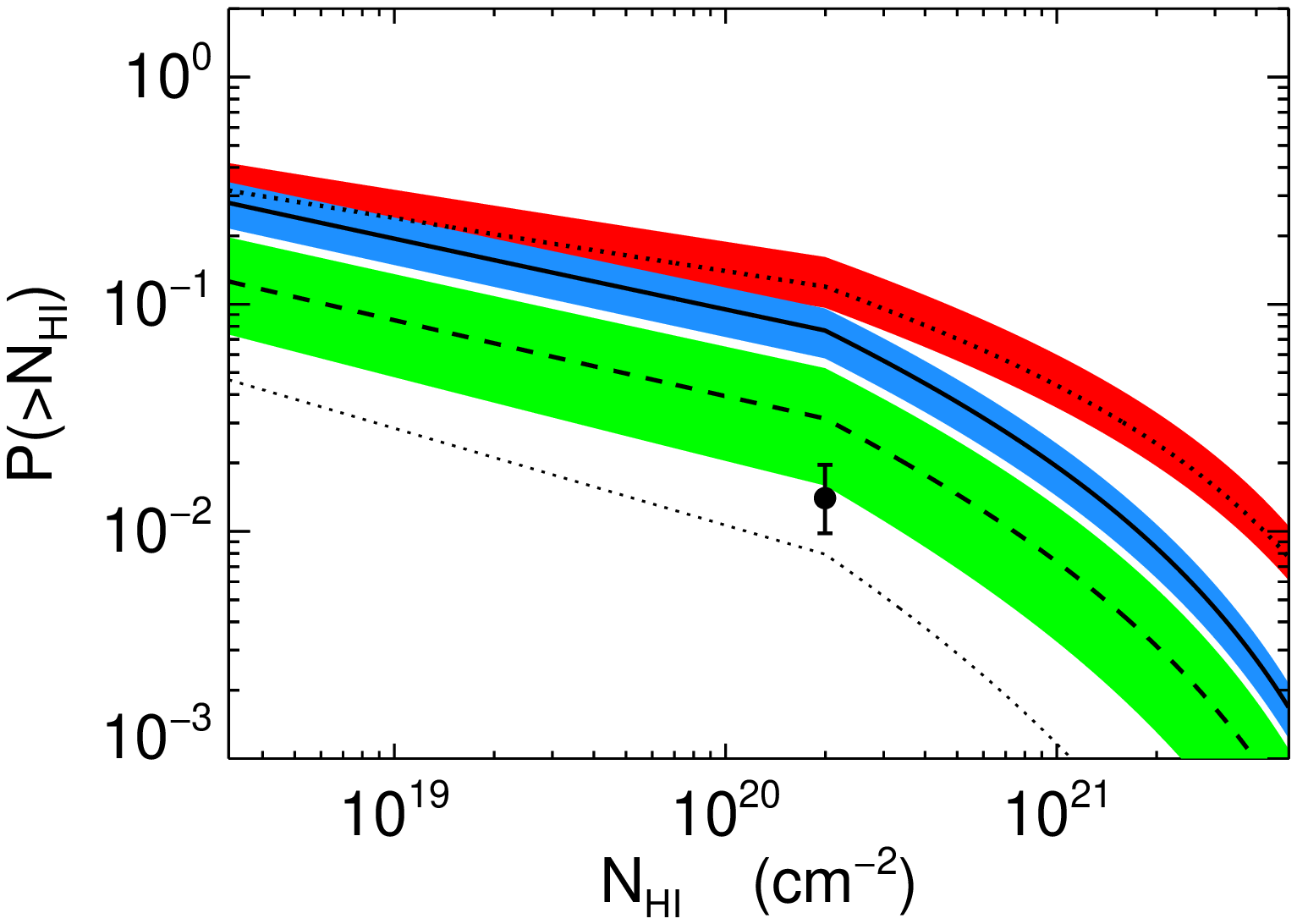,bb=82 72 576 412,
      width=0.5\textwidth}}
  \caption{ Probability of a quasar at $z\sim 2.5$ having 
    a proximate absorber within $3000~\kms$ as a function of limiting 
    column density.  The left panel is for `small' absorber cross sections 
    and the right panel is for `large' absorbers. Our measurement
    of the quasar-absorber correlation function $\xi_{\rm QA}$ from the 
    transverse direction is used to predict the number of proximate absorbers
    along the line-of-sight.   The curves and shaded regions show the 
    predictions and  $1\sigma$ errors for the quasar-absorber correlation 
    function $\xi_{\rm QA}$ estimated from the transverse direction in 
    \S~\ref{sec:cluster}, as well as the clustering strength measured by AS05. 
    The linestyles and colors correspond to the same models as in 
    Figure~\ref{fig:corr}. The lower dotted line indicates the prediction 
    in the absence of clustering, which is just the integral of the column
    density distribution in eqn.~\ref{eqn:fofn}. The point with the error
    bar represents the \citet{REB06} measurement for proximate DLAs 
    ($\log \mnhi > 20.3$) at $z\sim 3.4$.
    \label{fig:proximity}}
\end{figure*}

If the clustering pattern of optically thick absorbers around quasars
is isotropic, then we can use our estimate of the quasar-absorber
correlation function in \S~\ref{sec:cluster} to predict the number of
`proximate' super-LLSs that should be observed in a given velocity
window, $\Delta v$, along the line-of-sight. According to
eqn.~(\ref{eqn:pdla}), the clustering enhancement
$\chi_{\rm \parallel}$ can be computed by averaging $\xi_{\rm QA}$
over a cylinder of length $\Delta v \slash a H(z)$ and cross sectional
area $A$.

However, because the cross-section of the absorbers is unknown,
relating the transverse clustering to the line-of-sight clustering
requires an assumption about the size of the cross section.  Very
little is known about the sizes of DLAs and LLSs. \citet{Briggs89}
detected \ion{H}{1} 21cm absorption of a DLA with $\mnhi = 5\times
10^{21}~{\rm cm^{-2}}$ against an extended radio source, allowing them
to place a lower limit of $r \gtrsim 12~\hkpc$ on the size of the
absorbing region. \citet{Prochaska99} estimated an absorbing path
length of $\sim 2 \hkpc$ for a super-LLS with $\log \mnhi = 19.1$, by
comparing the column density to the total volume density, which was
determined from the collisionally excited \ion{C}{2}$^\ast$ $\lambda
1335$ transition. \citet{Adel06} constrained the size of a DLA with
$\log \mnhi = 20.4$ to be $\gtrsim 8~\hkpc$, based on their detection
of fluorescent Ly$\alpha$ emission from the edge of a DLA-galaxy
situated $\sim 8~\hkpc$ from the location of DLA
absorption. \citet{Lopez05} measured identical column densities of
$\mnhi = 10^{20.5}~{\rm cm}^{-2}$ for both DLAs at ($z=0.9313$)
detected in the individual images of the gravitational lens HE 0512
$-$ 3329, allowing them to constrain the size of the DLA to be
$\gtrsim 4\hkpc$ \citet[see also][]{Smette95}.  It is not
clear how to relate these measurements to the absorption cross
sections of the  $\log \mnhi > 19$ of interest to us here.

To determine the cross section size as a function of limiting column,
$r_{\rm abs}(>\mnhi)\equiv \sqrt{a A(>\mnhi)\slash \pi}$, we adopt the
simple approximation that the comoving number density of absorption
line systems, $n$, and the covering factor, $f_{\rm cov}$ are
independent of column density, which gives the simple scaling $r_{\rm
  abs}\propto \sqrt{\langle dN\slash dz\rangle}$ from
eqn.~(\ref{eqn:dndz}). Two families of sizes, `small' and `large' are
considered, which we believe bracket the range of
possibilities. Fiducial physical sizes of $r_{\rm abs} = 5~\hkpc$ and $r_{\rm
  abs}=20~\hkpc$ are chosen at the DLA threshold ($\log \mnhi > 20.3$)
for the `small' and `large' absorbers, respectively. The $r_{\rm
  abs}\propto \sqrt{\langle dN\slash dz\rangle}$ scaling then predicts
respective sizes of $r_{\rm abs}= 9~\hkpc$ and $r_{\rm abs}= 38~\hkpc$
for super-LLSs with $\log \mnhi > 19$.

In Figure~\ref{fig:proximity} we show the probability that a quasar at
$z=2.5$ will have a `proximate' optically thick absorption line
systems within $\Delta v <3000~\kms$, as a function of limiting column
density. The left (right) panel shows the prediction for the `small'
(`large') family of cross-section sizes, and the dot-dashed curves
shows the prediction in the absence of clustering
($\chi_{\parallel}=0$), which is simply the integral of the column
density distribution in eqn.~(\ref{eqn:fofn}). Since we have measured
the transverse clustering only for $\log \mnhi > 19$, this figure assumes 
that clustering is independent of limiting column density. 

The transverse clustering overpredicts the fraction of quasars which
have a proximate absorber by a large factor. For example, for `small'
absorption cross sections and $\gamma=1.6$, the quasar-absorber
correlation function measured from the transverse direction predicts
that a fraction $P=0.30\pm 0.07$ of all quasars should show a
proximate super-LLS ($\log \mnhi > 19$) within $\Delta
v=3000~\kms$. Even the AS05 clustering amplitude, which is a factor of
$\sim 3$ smaller than our best fit, would predict
$P=0.12^{+0.08}_{-0.06}$. A steeper correlation function results in
even more proximate absorbers. For the best fit transverse clustering
amplitude with $\gamma=2$, the probability would be
$P=0.49^{+0.17}_{-0.10}$. 

Making the absorbers larger changes this prediction by factors of
$\sim 2-3$.  The `large' absorbers predict a fraction $P=0.14\pm 0.03$
of quasars should have a proximate absorber, given our best-fit
clustering amplitude for $\gamma=1.6$. Our steeper $\gamma=2$ fit
gives $P=0.14^{+0.02}_{-0.04}$ and the AS05 result predicts $P=0.07\pm
0.03$ and our steeper $\gamma=2$ fit gives $P=0.14^{+0.02}_{-0.04}$.
Although the line density of proximate super-LLSs near quasar has yet
to be measured \citep[but see][]{Prochter06}, it is incontrovertible
that $15-50\%$ percent of quasars do not show a super-LLS within
$\Delta v < 3000~\kms$ along the line of sight.

\citet{REB06} recently measured the line density of proximate DLAs
($\log \mnhi > 20.3$) with $\Delta v <3000~\kms$ from 731 SDSS
quasars. The median redshift of their quasar sample was $z =3.335$ and
they found $\langle dN\slash dz \rangle = 0.38^{+0.18}_{-0.08}$. The
point with the error bar in Figure~\ref{fig:proximity} indicates the
probability of having a proximate DLA from the \citet{REB06}
measurement, $P_{\rm DLA}=0.013^{+0.006}_{-0.002}$. Our best-fit
clustering amplitude predicts that $P=0.12\pm 0.3$
($P=0.25^{+0.09}_{-0.05}$) of quasars should have a nearby DLA for
small absorbers and $\gamma=1.6$ ($\gamma=2$). Large absorbers change
this prediction to $P=0.055\pm 0.013$ ($P=0.069^{+0.023}_{-0.013}$)
\footnote{Comparing the \citet{REB06} measurement with our prediction
  ignores evolution in the clustering and the line density between the
  mean redshift of our sample ($z=2.5$) and theirs ($z =3.335$). The
  line density changes by $\sim 50\%$ over this redshift range
  \citep{phw05}, but this is small compared to the large discrepancy
  with prediction of the transverse clustering.}.

% The \citet{REB06} could be consistent
%with the weaker quasar-LBG clustering measured by AS05 if the
%absorbers are large $P=0.024^{+0.014}_{-0.011}$; small absorbers are
%inconsistent with AS05 at $\sim 1\sigma$.

The caveat should be included that our prediction for the number of
proximate absorbers from the transverse clustering, is very sensitive
to the small scale behavior of the correlation function. Specifically,
using eqn.~\ref{eqn:pdla} to predict the line-of-sight clustering
implicitly assumes that the power law model of the correlation
function $\xi \propto r^{-\gamma}$, is valid down the lower limit of
the integration, $Z_{\rm cut}$, which we have set to be the diameter
of the absorbers.  For super-LLSs, we used (proper) radii of $r_{\rm
  abs} = 9~\hkpc$ and $r_{\rm abs} =38~\hkpc$, for small and large
absorbers, respectively, corresponding to (comoving) $Z_{\rm cut} =
63$ and $Z_{\rm cut}=266$ at $z=2.5$. The smallest transverse
separation of our projected pair sample is $R=72~\hkpc$, and there are
just 5 sightlines with $R\lesssim 300~\hkpc$, three of which have an
absorber (see Figure~\ref{fig:scatter} and
Table~\ref{table:slls}). The transverse measurement assumes a single
power law from $R=70~\hkpc-5~\hMpc$. One should bear in mind that only a
handful of projected pairs probe the small scales ($R\lesssim
300~\hkpc$) that the line-of-sight clustering is very sensitive too;
but, it is reassuring that closest bin in Figure~\ref{fig:corr} is
consistent with the power law fit.

We have shown that the transverse clustering which we quantified in
\S~\ref{sec:cluster} overpredicts the abundance of proximate absorbers
along the line-of-sight by a large factor $\sim 4-20$, under
reasonable assumptions about the sizes of the cross sections of these
absorbers. The clustering pattern of absorbers around quasars is thus
highly anisotropic. The most plausible explanation for this anisotropy
is that the transverse direction is less likely to be illuminated by
ionizing photons than the line-of-sight, and that the optically thick
absorbers along the line-of-sight are being photoevaporated. We will
discuss the physical effects which could give rise to this anisotropy
in \S~\ref{sec:conc}. Next, we introduce a simple model that provides
physical insight into the problem of optically thick absorbers subject
to the intense ionizing flux of a nearby quasar.

\section{Photoevaporation of Optically Thick Clouds}
\label{sec:photo}
\begin{figure*}
  \centerline{\epsfig{file=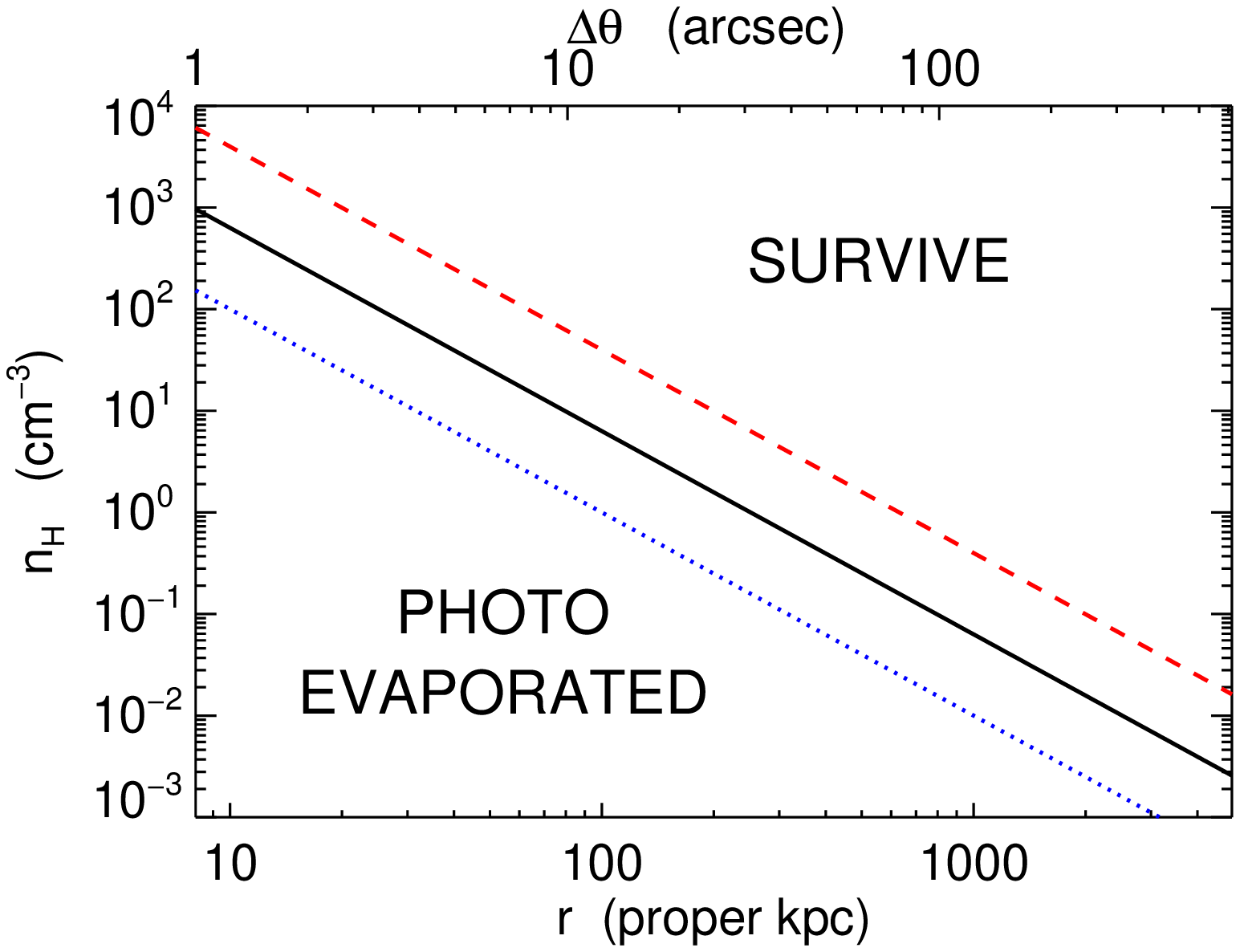,bb=80 50 560 420
      ,width=0.50\textwidth}
    \epsfig{file=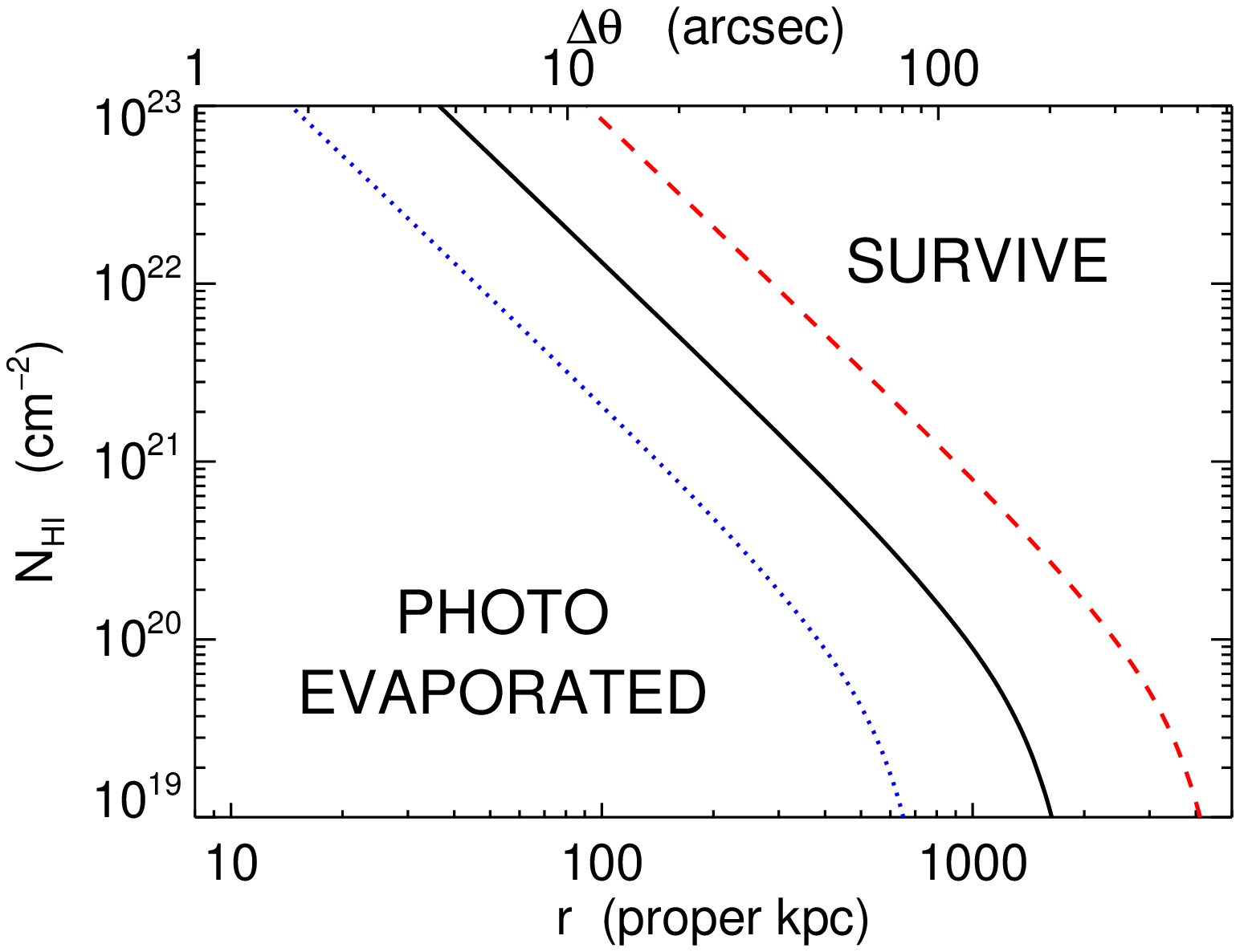,bb=80 50 560 420
      ,width=0.50\textwidth}}
  \caption{Regions of parameter space where an optically thick
    absorber can survive near a luminous quasar set by
    eqn.~\ref{eqn:delta}. The dashed (red), black (solid) and dotted
    (blue) curves correspond to $r=17$, $19$, \& $21$ mag,
    respectively. Optically thick absorbers above a curve can survive,
    whereas those below are photoevaporated. \emph{Left:} Survival
    regions in the volume density-distance plane, assuming a neutral
    column of $\log \mnhi = 20.3$. \emph{Right:} Survival regions in
    the neutral column density-distance plane, assuming a volume
    density of $n_{\rm H} =0.1~{\rm cm^{-3}}$.\label{fig:delta}}
\end{figure*}

The problem of an optically thick absorption line system exposed to
the ionizing flux of a nearby luminous quasar is analogous to that of
a neutral interstellar cloud being exposed to the ionizing radiation
of an OB star -- a problem which was first investigated by
\cite{OS55}. \citet{Bertoldi89} classified the behavior of
photoevaporating clouds based on their initial column density and the
ionization parameter at the location of the cloud, and developed an
analytical solution to follow the radiation-driven implosion phase of
the cloud. Interestingly, although \citet{Bertoldi89} was primarily
concerned with the fate of interstellar clouds near OB stars in
\ion{H}{2} regions, he commented briefly on the applicability of the same
formalism to Ly$\alpha$ clouds exposed to the ionizing flux of a quasar.

\subsection{Cloud Zapping}

Following \citet{Bertoldi89}, we model an optically thick absorber as
a homogeneous spherical neutral gas cloud with total number density of
hydrogen $n_H$, which is embedded in photoionized intergalactic medium
at temperature $T=20,000~{\rm K}$, corresponding to an isothermal
sound speed $c_{\rm i}=16.5~T_{20}^{1\slash 2}~\kms$, where $T_{20}$
is the temperature in units of $20,000~{\rm K}$. If the cloud is at a
distance $r$ from a luminous quasar which is emitting $S$ ionizing
photons per second, then the the ionizing flux at the cloud surface,
$F_{\rm i}=S\slash 4\pi r^2$, will drive an ionization front into the
neutral gas.  If the flux is large enough to make the conditions
R-type (see \citet{Spitzer} for a discussion of how ionization fronts
are classified), an R-type ionization front will propagate through the
cloud without dynamically perturbing the neutral gas. As the front
propagates into the cloud, it ionizes an increasing column of neutral
gas, steadily reducing the ionizing flux at the front, until the
conditions become M-type, at which point the front will stall and
drive a shock into the neutral upstream gas. This shock will implode
the cloud and compress it until conditions become D-type, allowing
the front to continue propagating and establishing a steady
photoevaporation flow \citep{Bertoldi89}.

The cloud will be `zapped', or completely photoevaporated, if the
ionizing flux is large enough relative to the column density, such
that the entire cloud can be ionized in a recombination time. In this
case, the R-type front will completely cross the cloud without
stalling.  Part of the cloud will remain neutral and a shock will form
provided that
\be 
\delta = 494 N_{\rm H,20.3}^{-1}\left[\Gamma-1.1\times 10^{-4}~
  T_{\rm 20}^{1\slash2}\right] < 1\label{eqn:delta}
\ee
where $N_{\rm H,20.3}$ is the \emph{total} hydrogen column in units of 
$10^{20.3}~{\rm cm^2}$, and the ionization 
parameter is defined by $\Gamma \equiv F_{i}\slash n_0 c$ with 
\be
\Gamma= 2.58\times 10^{-5}~S_{56}~r_{\rm Mpc}^{-2}~n_{\rm H,1}^{-1}, 
\ee
where $S_{56}$ is the ionizing flux in units of $10^{56}~{\rm s^{-1}}$, 
$r_{\rm Mpc}$ is the physical distance in units of ${\rm Mpc}$, and 
$n_{\rm H,1}$ is the total hydrogen number density in units of $1~{\rm cm^{-3}}$.
%JXP -- QSOs are harder than O and B stars.  Does this make a noticable 
%  difference in the calculation?
The condition that the ionization front be R-type at the surface of the
cloud is $\Gamma > 1.1\times 10^{-4}T_{20}^{1/2}$. 

Consider our fiducial example of a foreground quasar with $r=19$ at an
angular separation of $\Delta \theta = 1^{\prime}$ from an absorber
corresponding to a transverse proper distance of $485~{\rm kpc}$.  
The ionizing flux is enhanced by $g_{\rm UV}=130$ over the UV
background and $S_{56}=5.2$. For a DLA with \emph{total} hydrogen
column of $N_{\rm H}=10^{20.3}~{\rm cm^{-2}}$ at this distance, the
cloud will survive ($\delta < 1$) provided $n_{\rm H} > 0.27~{\rm
  cm^{-3}}$, which would give $\Gamma < 0.002$. 

The left panel of Figure~\ref{fig:delta} shows the lower limits on the
volume density of a DLA with neutral hydrogen column $\log \mnhi =
20.3$, set by the condition for cloud survival, as a function of
physical distance from a quasar at $z=2.5$. The dashed (red), black
(solid) and dotted (blue) curves correspond to $r=17$, $19$, \& $21$ mag,
respectively.  In the region below each curve, the volume densities
are too small and the clouds are photoevaporated; whereas, above the
curves the clouds can survive.  The right panel shows the lower limit
on the neutral hydrogen column density as a function of distance,
assuming a volume density $n_{\rm H}=0.1$.  If DLAs with $\log \mnhi =
20.3$ have sizes in the range $r_{\rm abs} \sim 1-5~{\rm kpc}$ their
corresponding number densities are $n_{\rm H} \sim N_{\rm H}/r_{\rm
  abs} = 0.01-1$. Thus according to Figure~\ref{fig:delta}
\emph{optically thick absorbers with $n_{\rm H} \lesssim 0.1$ will be
  photoevaporated if they lie within $\sim 1~{\rm Mpc}$ of a luminous
  quasar.}

Note that the condition in eqn.~(\ref{fig:delta}) considers the total
column density $\log N_{\rm H}$, not the neutral column, $\log
\mnhi$. Although DLAs are expected to be predominantly neutral $\mnhi
\approx N_{\rm H}$, the ionic absorption lines typically observed in
LLSs suggest that they are the photoionized analogs of DLAs
\citep{Prochaska99} and that their ionization state is determined by
ionization equilibrium with the UV background. In
eqn.~(\ref{eqn:delta}), we have assumed that optically thick absorbers
are spherical top-hat density distributions. Hence, an LLSs with
column density $\mnhi\lesssim 20$, can be thought of as a sightline
that passes through the photoionized outskirts of a top-hat cloud
which has a larger total hydrogen column $N_{\rm H}$, and that the
total column determines the ability of the cloud to survive a blast of
ionizing radiation from the quasar. Thus for $\mnhi\lesssim 20$, we
compute an ionization correction with the standard approach, which is
to assume a slab geometry and determine the ionization balance in a
uniform background, using a photoionization code such as Cloudy
\citep{CLOUDY}. To construct the curve in the right panel of
Figure~\ref{fig:delta}, we interpolated through a grid of Cloudy
solutions \footnote{The photoionization models used Cloudy version
  6.0.2 and used a Haardt \& Madau (2006, in preparation) QSO+galaxies
  spectrum at $z=2.5$ and assumed a metallicity of -1.5 and
  $J_{912}=-21.2$.} for $\mnhi(N_{\rm H})$.

\subsection{A Toy Model to Predict the Statistics of Proximate DLAs}
\label{sec:toy}

\begin{figure}
  \centerline{\epsfig{file=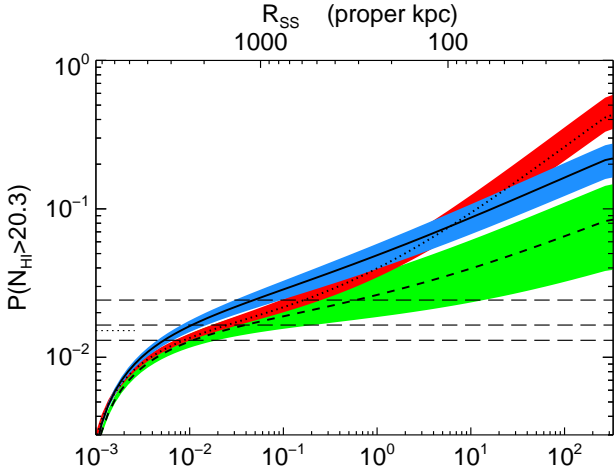,bb= 80 60 580 450,width=0.50
      \textwidth}}
  \caption{Toy model prediction for the probability of a quasar at $z
    \sim 3.4$ with $i=19.5$ having a proximate DLA within $3000~\kms$
    as a function of volume density.  The upper x-axis shows self-shielding
    radius, which is the minimum distance from the quasar at which a DLA 
    can survive for a given volume density. The curves and shaded regions
    show the predictions and $1\sigma$ errors for the quasar-absorber
    correlation function $\xi_{\rm QA}$ estimated from the transverse
    direction in \S~\ref{sec:cluster}, as well as the clustering
    strength measured by AS05.  The linestyles and colors correspond
    to the same models as in Figure~\ref{fig:corr}.  The long-dashed
    horizontal lines indicate the measurement and $\pm 1\sigma$ range
    measured by \citet{REB06} $P_{\rm DLA}=0.013^{+0.006}_{-0.002}$
    . The short dotted horizontal line is the cosmic average (i.e. no
    clustering) measured by \citep{phw05}.  This model suggests that
    DLAs with volume densities in the range $10^{-2}~{\rm
      cm^{-3}}\lesssim n_{\rm H}\lesssim 10^{-1}~{\rm cm^{-3}}$ are
    required to agree with the proximate DLA measurement of
    \citet{REB06}.\label{fig:toy}}
\end{figure}

In this section we use a toy model to illustrate how quasar-absorber
clustering can be used to constrain the physical properties of
optically thick absorbers.  The criteria $\delta < 1$ in
eqn.~(\ref{eqn:delta}), gives a minimum distance from the quasar,
$R_{\rm SS}(n_{\rm H})$ as a function of volume density, at which an
optically thick absorption line system with column density $\mnhi$ can
survive. Our approach is to simply assume that absorbers at smaller
distances are photoevaporated and absorbers at larger distances
survive. We also assume that the transverse direction is not
illuminated by the quasar, and hence the transverse clustering
measures the \emph{intrinsic} quasar-absorber clustering, in the
absence of ionization effects.  Because proximate absorbers are
definitely illuminated, this intrinsic clustering is then reduced along the
line-of-sight by photoevaporation.

For the column density range $\log \mnhi > 20.3$, we evaluate
$R_{\rm SS}$ at the lower limit $\log \mnhi = 20.3$, which is a decent
approximation because the column density distribution is steep, and
the statistics will be dominated by absorbers near the threshold.  To
simplify the computation, we take $R_{\rm SS}$ to be a distance only along
the line of sight, which is valid provided that $R_{\rm SS} \gg
\sqrt{A}$.  Then we can write that the line density of
absorbers within $\Delta v < 3000~\kms$ is
\be 
\frac{dN}{dz}=
\left\langle\frac{dN}{dz}\right\rangle \left[1 - \frac{a R_{\rm
      SS}H(z)}{\Delta v} + \chi_{\parallel}(\Delta v)\right],
\label{eqn:toy}
\ee
where $\chi_{\parallel}$ is given by eqn.~(\ref{eqn:pdla}) but with
$Z_{\rm cut} = R_{\rm SS}$. The assumption that the transverse direction gives
the intrinsic clustering in the absence of ionization effects amounts to 
using the correlation function $\xi_{\rm QA}$, measured from the transverse 
clustering, in the line-of-sight integral in eqn.~(\ref{eqn:pdla}). 

In Figure~\ref{fig:toy} we show our toy model prediction for the
probability of a quasar having a proximate DLA with $\log \mnhi > 20.3$
within $\Delta v < 3000~\kms$, as a function of total volume density
of hydrogen $n_{\rm H}$. We assumed that the DLAs have a size of
$r_{\rm abs} = 5~\hkpc$ characteristic of the `small' absorbers
discussed in \S~\ref{sec:proximity}. However, identical results are
obtained for `large' absorbers ($r_{\rm abs} = 20~\hkpc$). Because our toy 
model excludes small scales $r_{\rm abs} \ll R_{\rm SS}$ from
the clustering integral in eqn.~(\ref{eqn:pdla}), we are again in the
far-field limit where the the volume average is nearly independent of
the absorber cross section.  This independence breaks down for very
large densities $n_{\rm H} \gtrsim 10^2$, where $R_{\rm SS}$
approaches the size of the absorbers (see Figure~\ref{fig:toy}).
The curves and shaded regions show the predictions and $1\sigma$
errors for the transverse clustering measured in \S~\ref{sec:cluster},
as well as the clustering strength measured by AS05. We assumed a quasar
with magnitude $i=19.5$ at $z=3.335$, chosen to match the mean
magnitude and redshift of the proximate DLA sample of
\citet{REB06}.  The long-dashed horizontal lines indicate
the measurement and $\pm 1\sigma$ range measured by \citet{REB06},
$P_{\rm DLA}=0.013^{+0.006}_{-0.002}$ . 

Our toy model suggests that DLAs with volume densities in the range
$10^{-2}~{\rm cm^{-3}}\lesssim n_{\rm H}\lesssim 10^{-1}~{\rm
  cm^{-3}}$ are required to agree with the measurement of
\citet{REB06}. Even the weaker AS05 clustering of LBGs around quasars
would require $n_{\rm H}\lesssim 1~{\rm cm^{-3}}$.  For larger volume
densities, DLAs illuminated by quasars would survive at smaller radii
where the clustering is strong, giving rise to a larger number of
proximate DLAs. Note that the comparison of our toy model to the
\citet{REB06} measurement in Figure~\ref{fig:toy} assumes that the
clustering of absorbers is independent of column density threshold
(i.e.  our measurement is for $\log \mnhi > 19$) and that there is
negligible evolution in the clustering between the mean redshift of
our clustering sample, $z=2.5$ and that of the \citet{REB06} sample
$z=3.335$. These assumptions were made because our quasar-absorber
sample did not have sufficient statistics to measure the clustering
(see \S~\ref{sec:cluster}) in the same redshift and column density
range as the \citet{REB06} proximate DLA measurement.

Although crude, this model illustrates how a comparison of
line-of-sight and transverse quasar absorber clustering can be used to
determine the density distribution in optically thick
absorbers. Detailed models of self-shielding with radiative transfer
\citep{ZM02a,Cantalupo05,Juna06} would be required for a more accurate
treatment, and such analyses could easily include cuspy density
profiles or a multiphase distribution of gas. Although better
statistics and more theoretical work are necessary, the clustering
of optically thick absorbers around quasars will provide important new
constraints on the physical nature of these systems.

\section{Summary and Discussion}
\label{sec:conc}

\subsection{Summary}

In this paper we used a sample of 17 super-LLSs ($\log \mnhi < 19$)
selected from 149 projected quasar pairs sightlines in QPQ1 to
investigate the clustering pattern of optically thick absorbers around
quasars. Based on this data, this paper presented the following
results:

\begin{enumerate}

\item A simple formalism was presented for quantifying the clustering
  of absorbers around quasars in both the transverse and line-of-sight
  directions. The clustering of absorbers around quasars in the
  transverse direction is independent of the size of the absorption
  cross section; whereas, the line-of-sight clustering was shown to be
  sensitive to the cross section size.  

\item Applying this formalism to the 17 super-LLSs ($\log \mnhi < 19$)
  selected from 149 projected quasar pair sightlines with mean
  redshift $z=2.5$, we determined a comoving correlation length of
  $r_0=9.2^{+1.5}_{-1.7}~\hMpc$ for a power law correlation function
  with $\gamma=1.6$. This is three times stronger than the clustering
  of LBGs around quasars recently measured by AS05.  If we assume a
  steeper slope of $\gamma=2.0$, we measure $r_0=5.8^{+1.0}_{-0.6}~\hMpc$.

\item The clustering of optically thick absorbers around quasars is
  highly anisotropic. If we apply the clustering amplitude measured in
  the transverse direction to the line-of-sight, the fraction of
  quasars which have a proximate absorber within $\Delta v <
  3000~\kms$ is overpredicted by a factor as large as $\sim 4-20$, depending
  on assumptions about cross section sizes and the slope of
  the correlation function. The most plausible explanation for the
  anisotropy is that the transverse direction is less likely to be
  illuminated by ionizing photons than the line-of-sight, and that the
  optically thick absorbers along the line-of-sight are being
  photoevaporated.  

%JXP -- Advertise the MgII paper on this point too?

\item A simple model of absorbers as uniform spherical overdensities
  is discussed and we write down an analytic criterion which
  determines whether an absorber illuminated by a quasar will be able
  to self-shield. This criterion indicates that optically thick
  absorbers with $n_{\rm H} \lesssim 0.1$ will be photoevaporated if
  they lie within $\sim 1~{\rm Mpc}$ of a luminous quasar.  We combine
  this criterion with a toy model of how photoevaporation affects the
  line-of-sight clustering, to illustrate how comparisons of the
  line-of-sight and transverse clustering around quasars can
  ultimately be used to constrain the distribution of gas in optically
  thick absorption line systems.

\end{enumerate}

\subsection{Discussion}

The anisotropic clustering pattern of absorbers around quasars
suggests that the transverse direction is less likely to be illuminated
by ionizing photons than the line-of-sight.  This suggestion gains
credibility in light of the  null detections of the transverse
proximity effect in the Ly$\alpha$ forests of projected quasar pairs
\citep[][but see Jakobsen \etal
2003]{Crotts89,DB91,FS95,LS01,Schirber04,Croft04}.  Although these
studies are each based only on a handful of projected pairs, they all
come to similar conclusions: the amount of (optically thin) Ly$\alpha$
forest absorption, in the background quasar sightline near the
redshift of the foreground quasar, is \emph{larger} than average
rather than smaller -- the opposite of what is expected from the
transverse proximity effect. Two physical effects can explain both the
optically thin results and our result for optically thick systems:
anisotropic emission or variability, which we discuss in turn.

If quasar emission is highly anisotropic, the line-of-sight would be
exposed to the ionizing flux of the quasar; whereas, transverse
absorbers would be more likely to lie in shadowed regions.  Studies of
Type II quasars and the X-ray background suggest that quasars with
luminosities comparable to our foreground quasar sample ($M_{\rm B} <
-23$) have $\sim 30\%$ of the solid angle obscured
\citep{Ueda03,Barger05,TU05}, although these estimates are highly
uncertain. Naively, we would expect the covering factor of transverse
absorbers to be approximately equal to the average fraction of the
solid angle obscured. But in QPQ1 we found a very high covering factor
(6/8) for having an optically thick absorber with $\log \mnhi > 17.2$
(see Figure~1 of QPQ1) on the smallest (proper) scales $R <
150~\hkpc$. Although the statistics are clearly very poor, this high
covering factor is suggestive of a significantly larger obscured
fraction.

If the ionizing flux of the foreground quasar varies considerably on a
timescale shorter than the transverse light crossing time between the
foreground and background sightlines, a transverse proximity effect
might not be observable. This is because the ionization state of the
gas along the transverse sightlines is sensitive to the foreground
quasars luminosity a light crossing time \emph{before} the light that
we observe was emitted. At $60\arcsec$ ($1.2~\hMpc$) from a $z=2.5$
quasar the transverse light crossing time is $1.1\times
10^6$~yr. 

Currently, the lower limit on the intermittency of quasar emission
comes from observations of the (optically thin) proximity effect
\citep{BDO88,Scott00}, in the Ly$\alpha$ forests near quasars.  The
presence of an optically thin proximity effect implies that the IGM
has had time to reach ionization equilibrium with the quasars
increased ionizing flux, which requires that the duration of a burst
of quasar radiation is longer than the IGM equilibration time, $t_{\rm
  burst} \gtrsim 10^4$~yr \citep{Martini04}. The photoevaporation
timescale for an optically thick absorber is $\sim \mnhi\slash F$ or
the light crossing time, whichever is longer, where $F$ is the
ionizing flux. At a distance of $100~{\rm kpc}$ from an $i=19$ quasar,
it would take $15,000$ yr to photoevaporate a DLA with $\log \mnhi =
20.3$, i.e. comparable to the light crossing time if DLAs have sizes
of $\sim 5~{\rm kpc}$. Thus the time it takes for the line-of-sight to
manifest the effects of the quasars ionizing flux is $\sim 10^4~{\rm
  yr}$ for both the optically thin and optically thick regimes. Hence,
if quasars emit in bursts of duration $10^4~{\rm yr}\lesssim t_{\rm
  burst} \lesssim 10^6~{\rm yr}$, this would suffice to explain the
absence of the optically thin transverse proximity effect as well as
the anisotropic clustering pattern of optically thick absorbers
discovered here.

The optically thin proximity effects are probably the most
promising way to disentangle whether obscuration or variability can
explain why the transverse direction is less likely to be illuminated
by ionizing photons \citep{Schirber04,Croft04}. This is because for
the lower column density absorbers characteristic of the
Ly$\alpha$ forest, cosmological N-body simulations can predict the
statistical properties of the absorbers \emph{ab initio} without the
complications of radiative transfer.

It might seem puzzling that we measure a clustering amplitude which is
a factor of three larger than the clustering of LBGs around quasars at
similar redshift measured by AS05. One could argue that on small
scales we are probing material intrinsic to the quasar.  If
interpreted as clustering, this intrinsic contribution would lead us
to overestimate the correlation function. However, only our closest
sightline ($R = 22~\hkpc$, proper) is sufficiently small for this
to be a worry, and this would imply a very large cross section for DLA
absorption. AS05 excluded small scales from their analysis, whereas,
the strength of our signal is in part driven by the small scale
systems. If we exclude sightlines with $R < 500~\hkpc$ we measure a
$r_0=7.6^{+1.9}_{-2.0}\hMpc$ which is within $1\sigma$ of the AS05
measurement. It is also possible that we measured a larger clustering
signal because the galaxies which host optically thick absorbers are
more strongly biased with respect to the dark matter distribution than
LBGs \citep[but see][]{Cooke06}.

We argued that the transverse clustering predicts that
galaxies correlated with a quasar should give rise to an absorption
line systems with $\log \mnhi > 19$ within $\Delta v < 3000~\kms$ a
significant fraction of the time $\sim 15-50\%$. We postulated that
these systems are not observed because of photoevaporation, but
shouldn't a similar argument also apply to galaxies?  High redshift
galaxy spectra should show optically thick absorption due to nearby
correlated galaxies, in addition to intrinsic absorption due to
\ion{H}{1} in the galaxy itself. The individual spectra of high
redshift galaxies rarely have high enough signal-to-noise ratio to
detect even DLAs.  However, the break due to Lyman limit absorption
can be detected in very deep integrations with the added advantage
that the LLSs are much more abundant. Recently, \citet{Shapley06}
detected emission blueward of the Lyman limit in 2 out of 14 LBGs at
$z\sim 3$. For a galaxy-absorber correlation length in the range
$r_0=5-10~\hMpc$, eqn.~(\ref{eqn:clust}) would predict that $25-75\%$
of galaxies should show a correlated absorber within a window of
$\Delta v < 500~\kms$, comparable to the redshift uncertainties of
LBGS.  This number assumes LLSs with $\log \mnhi > 17.2$ have a size
of $\sim 50~\hkpc$ and we used the abundance of LLSs measured by
\citet{Peroux03}.  While it is possible that the lack of flux blueward
of the Lyman limit in LBGs is due to absorption by gas intrinsic to
the LBG, correlated absorbers along the line of sight could also play
a significant role.

Using the ionizing flux of a quasar to study the distribution of
neutral gas in optically thick absorbers is a powerful new way to
study these absorption line systems.  We showed how comparisons of the
line-of-sight and transverse clustering of absorbers around quasars
constrains their distribution of gas.  Theoretical models of LLSs and
DLAs which include radiative transfer and self-shielding
\citep{ZM02a,Cantalupo05,Juna06} should explore how morphology, cuspy
density profiles, or a multiphase medium, change the ability of
absorbers to survive near quasars.

Optically thick \ion{H}{1} clouds in ionization equilibrium with a
radiation field, re-emit $\sim 60\%$ of the ionizing photons they
absorb as fluorescent Ly$\alpha$ recombination line photons
\citep{GW96,ZM02b,Cantalupo05,Adel06,Juna06}. Recently,
\citet{Adel06} reported a detection of fluorescence from a
serendipitously discovered DLA ($\log \mnhi = 20.4$) situated
$49\arcsec$ away from a luminous quasar at $z=2.84$. Fluorescent
Ly$\alpha$ emission from absorbers illuminated by quasars offers a
new window to study physical properties of absorbers, as well as a
second transverse sightline to observe the ionizing flux of a quasar.
In QPQ1 we published seven new quasar-absorber pairs which would have
fluorescent surface brightnesses of $\mu_{{\rm
    Ly}\alpha}=19.5-24.3~{\rm mag~arcsec}^{-2}$, if the foreground
quasars emit isotropically. However, the clustering anisotropy
discussed here suggests that these transverse absorbers may not be
illuminated, and the lack of a detection of fluorescent emission would
provide compelling evidence in favor of this conclusion. Constraints
on fluorescent emission from these systems will be discussed in the
next paper in this series \citep{LLS3}. It is particularly intriguing
that proximate DLAs observed along the line-of-sight, seem to
preferentially exhibit \lya\ emission superimposed on the \lya\
absorption trough \citep{Moller98,Ellison02}.  Is this \lya\ emission
fluorescence?  Is the fluorescent emission from proximate DLAs
observable?  

%Theoretical work is required to determine the
%observational signatures of this emission, and what the emission
%profiles can teach us about the sizes, morphologies, densities, and
%kinematics of the illuminated absorbers \citep[see
%e.g.][]{ZM02b,Cantalupo05,Juna06}.

In QPQ1 we argued that a factor of $\sim 20$ more transverse
quasar-absorber pairs could be compiled in a modest amount of
observing time, which would improve our measurement of the transverse
correlation function by a factor of $\sim 5$. The study of
\citet{REB06} searched $\sim 700$ SDSS quasars for DLAs and found 12
proximate DLAs with $\Delta v < 3000~\kms$. However, the full SDSS
quasar survey has $\sim 10,000$ quasars with $z\gtrsim 2.2$
\citep{Schneider06}, which could be used to search for proximate
systems and increase the number known to $\gtrsim 100$. This would be
sufficient to measure the column density distribution $f(N_{\mnhi})$
of absorbers near quasars, and any differences between it and the
distribution in the `field' would provide an important new constraint
on the physical nature of optically thick absorbers.  Finally, we note
that similar studies using projected quasar pairs and searching for
line-of-sight proximate absorbers can be conducted for \ion{Mg}{2},
\ion{C}{4} or other metal absorption lines systems near quasars
\citep{Bowen06,Prochter06}, yielding similar insights into their
physical nature.

Large samples of optically thick absorption line systems near quasars
are well within reach, for transverse systems using quasar pairs as
well as proximate absorbers along the line-of-sight. This data will
provide new opportunities to characterize the environments of quasars,
the physical nature of absorption line systems, and it will uncover new
laboratories for studying fluorescent emission from optically thick
absorbers. 

%However, a significant amount of theoretical work still
%needs to be done before we can correctly interpret these observations.

% Models of extragalactic UV background usually include
%a significant contribution from galaxies, assuming some fraction of these
%photthat a significant contribution 
%that Lyman limit systems have a

\acknowledgements We are grateful to Jordi Miralda-Escud{\'e}, Juna
Kollmeier, Piero Madau, and Zheng Zheng for reading an early version of 
this manuscript and providing critical comments. 
We thank John O'Meara for helpful discussions and for sharing results 
prior to publication. JFH acknowledges enlightening discussions with Kurt
Adelberger, Doron Chelouche, Bruce Draine, Sara Ellison, Taotao Fang, 
Chris McKee, Brice Menard, David Russell, Alice Shapley, and Michael Strauss.

JFH is supported by NASA through Hubble Fellowship grant \# 01172.01-A awarded 
by the Space Telescope Science Institute, which is operated by the
Association of Universities for Research in Astronomy, Inc., for NASA,
under contract NAS 5-26555. JXP wishes to acknowledge funding through
NSF grant AST-0307408.

The conclusions of this work are based on data collected from
observatories at the summit of Mauna Kea. The authors wish to
recognize and acknowledge the very significant cultural role and
reverence that the summit of Mauna Kea has always had within the
indigenous Hawaiian community.  We are most fortunate to have the
opportunity to conduct observations from this mountain.

%\begin{appendix}
%
%  \section{Ionization Fronts}
%  
%  \citep{Spitzer} classifies ionization fronts through the isothermal
%  sound speed in the neutral gas, $c_0$, and the in the ionized
%  gas. Ionization fronts with $F_i < u_{\rm D} n_0$, where $u_{\rm
%    D}\simeq c_0^2\slash 2 c_i$, are called D-type and propagate
%  subsonically through the cloud. In the range $u_{\rm D} < F_{\rm
%    i}\slash n_0 < 2 c_i$, no solution to the ionization jump
%  conditions exists, and these fronts are classified as M-type. An
%  M-type ionization front will drive a shock into the neutral gas
%  cloud, compressing it until the density is sufficiently high to make
%  the shock D-type, such that the shock can propagate. Ionization
%  fronts with $F_i > 2 n_0 c_i$ are called R-type, and propagate
%  through the cloud without dynamically perturbing it.

%\end{appendix}

\end{document}